\documentclass[fleqn,usenatbib]{mnras}

\usepackage{newtxtext,newtxmath}

\usepackage[T1]{fontenc}

\DeclareRobustCommand{\VAN}[3]{#2}
\let\VANthebibliography\thebibliography
\def\thebibliography{\DeclareRobustCommand{\VAN}[3]{##3}\VANthebibliography}

\usepackage{graphicx}	
\usepackage{amsmath}	
\usepackage{subfig}

\newcommand{\galfit}{{\tt{GALFIT}}}

\title[High-z AGN]{EPOCHS VII: Discovery of high redshift ($6.5 < z < 12$) AGN candidates in JWST ERO and PEARLS data}

\author[Juodžbalis et al.]{
Ignas Juod\v{z}balis,$^{1}$\thanks{E-mail: ignas.juodzbalis@gmail.com}
Christopher J. Conselice,$^{1}$
Maitrayee Singh,$^{1}$
Nathan Adams,$^{1}$
\newauthor
Katherine Ormerod,$^{1}$
Thomas Harvey,$^{1}$
Duncan Austin,$^{1}$
Marta Volonteri,$^{2}$
Seth H. Cohen, $^{3}$
\newauthor
Rolf A. Jansen, $^{3}$
Jake Summers, $^{3}$
Rogier A. Windhorst, $^{3}$
Jordan C. J. D'Silva, $^{4, 5}$
Anton M. Koekemoer, $^{6}$
\newauthor
Dan Coe, $^{6, 7, 8}$
Simon P. Driver, $^{4}$
Brenda Frye, $^{9}$
Norman A. Grogin, $^{6}$
Madeline A. Marshall, $^{10, 5}$
\newauthor
Mario Nonino, $^{11}$
Nor Pirzkal, $^{6}$
Aaron Robotham, $^{4}$
Russell E. Ryan, Jr., $^{6}$
\newauthor
Rafael {Ortiz~III}, $^{3}$
Scott Tompkins, $^{3}$
Christopher N. A. Willmer, $^{9}$
Haojing Yan $^{12}$
\\
$^{1}$ Jodrell Bank Centre for Astrophysics, University of Manchester, Oxford Road, Manchester UK\\
$^{2}$ Institut d’Astrophysique de Paris, Sorbonne Universit\'e, CNRS, UMR 7095, 98 bis bd Arago, F-75014 Paris, France, \\
$^{3}$ School of Earth and Space Exploration, Arizona State University,
Tempe, AZ 85287-1404, USA \\
$^{4}$ International Centre for Radio Astronomy Research (ICRAR) and the
International Space Centre (ISC), \\ The University of Western Australia, M468, 35 Stirling Highway, Crawley, WA 6009, Australia \\ 
$^{5}$ ARC Centre of Excellence for All Sky Astrophysics in 3 Dimensions
(ASTRO 3D), Australia \\
$^{6}$ Space Telescope Science Institute, 3700 San Martin Drive, Baltimore, MD 21218, USA \\
$^{7}$ Association of Universities for Research in Astronomy (AURA) for the European Space Agency (ESA), STScI, Baltimore, MD 21218, USA \\
$^{8}$ Center for Astrophysical Sciences, Department of Physics and Astronomy, The Johns Hopkins University, 3400 N Charles St. Baltimore, MD 21218, USA \\
$^{9}$ Steward Observatory, University of Arizona, 933 N Cherry Ave,
Tucson, AZ, 85721-0009, USA \\
$^{10}$ National Research Council of Canada, Herzberg Astronomy \&
Astrophysics Research Centre, 5071 West Saanich Road, Victoria, BC V9E 2E7,
Canada \\
$^{11}$ INAF-Osservatorio Astronomico di Trieste, Via Bazzoni 2, 34124
Trieste, Italy \\
$^{12}$ Department of Physics and Astronomy, University of Missouri,
Columbia, MO 65211, USA
}

\date{Accepted XXX. Received YYY; in original form ZZZ}

\pubyear{2023}

\begin{document}
\label{firstpage}
\pagerange{\pageref{firstpage}--\pageref{lastpage}}
\maketitle

\begin{abstract}
We present an analysis of a sample of robust high-redshift galaxies selected from the `blank' fields of the Prime Extragalactic Areas for Reionization Science (PEARLS) survey and Early Release Observations (ERO) data from JWST with the aim of selecting candidate high-redshift active galactic nuclei (AGN). Sources were identified from this parent sample using a threefold selection procedure, which includes spectral energy distribution (SED) fitting to identify sources that are best fitted by AGN SED templates, a further selection based on the relative performance of AGN and non-AGN models, and finally morphological fitting to identify compact sources of emission, resulting in a purity-oriented procedure. Using this procedure, we identify a sample of nine AGN candidates at $6.5 < z < 12$, from which we constrain their physical properties as well as measure a lower bound on the AGN fraction in this redshift range of $5 \pm 1$\%.  As this is an extreme lower limit due to our focus on purity and our SEDs being calibrated for unobscured Type 1 AGN, this demonstrates that AGN are perhaps quite common at this early epoch.  The rest-frame UV colors of our candidate objects suggest that these systems are potentially candidate obese black hole galaxies (OBG).  We also investigate Chandra and VLA maps of these areas from which we calculate detection limits.   Of note is a $z = 11.9$ candidate source exhibiting an abrupt morphological shift in the reddest band as compared to bluer bands, indicating a potential merger or an unusually strong outflow.

\end{abstract}




\section{Introduction}

The origin and evolution of supermassive black holes remains an active area of research in astrophysics.  One of the major problems is that predicted masses for black hole seeds, which are expected to form from population III stars at $z=10-50$ \citep{first_pop3}, are too low to grow to the sizes observed at lower redshifts. For example, $z = 7.5$ quasars host black holes with masses in excess of $10^9 M_\odot$ \citep{z75_blackhole}, which are difficult to form unless super-Eddington accretion takes place \citep{stellarbh_mnras}.  Other formation models, such as the direct collapse of gas clouds, stellar collisions in dense clusters, and collapsing primordial density fluctuations, similarly lack conclusive observational evidence to distinguish them from one another as summarized in \cite{Volonteri2021}. 

A few candidate direct collapse black holes have been identified to date, pre-JWST.   This includes candidates identified by \cite{2016MNRAS_directcollapse} in the GOODS-S region of the CANDELS survey \citep{2011ApJS..197...35G,2011ApJS..197...36K}, which are very faint objects.  Thus, a next generation instrument, with increased survey depth is likely to identify more of such candidates (Nabizadeh et al. 2023 in prep.). The recent launch of the James Webb Space Telescope (JWST) has given us such an instrument and presents an excellent opportunity to start investigating the formation of young central massive black holes and start testing the validity of current black hole formation models by direct observations. While most black hole seeds are expected to have formed between redshifts of $z=14$ and $z=30$ \citep{mnras_collapse_era, ApJ_smbh}, which lies somewhat beyond the expected capabilities of the telescope, JWST may be able to detect accreting black hole seeds up to $z=10 - 12$ \citep{JWST_bhsearch}. It is also important to note that some black hole seed formation models, for instance \cite{Bellovary_2011} and \cite{late_pop_3} predict their formation to occur, albeit at a significantly diminished pace up to the end of reionization at $z \sim 5$, giving further credence to the idea that JWST surveys may be able to validate our current understanding on the origin of these objects.

Currently JWST efforts in tracing the evolution of galaxies have been focused largely on the detection of high redshift star forming galaxies (\cite{2023MNRAS.518.4755A}, \cite{2023MNRAS.518L..19R}, \cite{2022arXiv221105792F} and \cite{2022ApJ...940L..14N}) and morphological evolution of galaxies (\cite{2022ApJ...938L...2F}, \cite{2022arXiv220713527T} and \cite{2022ApJ...938L..24F}). The first year of observation also yielded three spectroscopically confirmed active galactic nuclei (AGN). Two at $z\approx5$ \citep{2022arXiv220907325O, hidden_monsters} and one at $z\approx8$ by \cite{CEERS_z8_agn}, as well as two candidates, one at $z\approx12$ by \cite{2022ApJ...938L..24F} and one at $z \approx 11$ by \cite{GZ-z11}. More recent studies by \cite{furtak2023jwst, LabbeAGN, matthee2023little} also hint at the presence of a significant population of partially obscured AGN at $z < 7$. However, no search aimed explicitly at searching for high redshift ($z > 7$) AGN candidates has been carried out in detail using imaging - a combination of both spectral energy distributions and morphology/structure.  

Furthermore, finding forming black holes through AGN is an important process that needs to be done photometrically, if possible, to find objects that can be followed up with Near Infrared Spectrograph (NIRSpec) spectroscopy.   This will be critical for determining the demographics of early AGN as well as 'naked' black holes that might exist at early times.  If these systems can be found photometrically, in principle, this would allow for large samples and demographics of this population to be characterised.    

In this work, we present the results for such a search for candidate AGN sources using JWST \emph{Near Infrared Camera} (NIRCam) imaging data \citep{NirCam_paper}.    We identify a small sample of excellent candidates for being high redshift AGN. We discuss in this paper our method for finding these objects and give a description of the properties of these potential early/young AGN or black holes, and provide a pathway to use our methods to find further larger samples of such objects. 

This paper is organised as follows - section \ref{sect:sel} contains a brief description of the data and the reduction process. Section \ref{sec:cand} presents a discussion of AGN identification methods used and checks of their validity, section  \ref{sect:src} presents an overview of the properties of the selected sources. The findings of this paper are summarized in section \ref{sec:conc}. Where appropriate we adopt a standard cosmology of $\Omega_m = 0.3$, $\Omega_\Lambda = 0.7$ and $H_0 = 70$ km s$^{-1}$Mpc$^{-1}$, all reported magnitudes use the AB system.

\section{Data}\label{sect:sel}

\subsection{Sample and Data reduction}

The galaxy sample from which we carry out this analysis derives from the Early Release Observations alongside the PEARLS GTO Survey fields: El Gordo, MACS-0416 and the North Ecliptic Pole Time Domain Field (NEP-TDF) \citep[][]{Windhorst2023}.  This data set is from the EPOCHS sample which are, a re-redcution and analysis of these fields to create a homogeneous reduction and ultimately a catalog, see \cite{UV_LF_methods} and Conselice et al. 2023, in prep.  This sample results from processing these data ourselves at all steps using our own refined methods that maximises our detection of faint galaxies and the accuracy of the photometry.  This paper is part VII in this series, with other forthcoming papers on the star formation rate, stellar populations, morphologies and stellar masses of this sample.  This paper is our first look at finding AGN within distant galaxies and can been as a first look at how to approach this problem.

Specifically, the NIRCam data used for source detection and photometry originated from the CEERS \citep[PID: 1345, PI: S. Finkelstein, see also][]{Bagley2022}, GLASS \citep[PID: 1324, PI: T. Treu,][]{Treu2022}, SMACS 0723 \citep[PID: 2736, PI: K. Pontoppidan,][]{Pontoppidan2022} and NGDEEP \citep[PID: 2079, PIs: S.\@ Finkelstein, Papovich and Pirzkal,][]{Bagley2023} public surveys. We also include GTO targets of El-Gordo, MACS 0416 and NEP from the Prime Extragalactic Areas for Reionization Science (PEARLS) survey (PI: R. Windhorst \& H.Hammel, PID: 1176 \& 2738, \citep{Windhorst2023}).

NIRCam filter sets used by these surveys were largely similar, with all of them utilizing some combination of the seven wide (F090W, F115W, F150W, F200W, F277W, F356W and F444W) and one medium width (F410M) filter.

The reduction procedure applied to all unprocessed JWST data products is described in detail by \cite{UV_LF_methods} and in Conselice et al. (2023, in prep), and can be summarized as follows: (1) Initial processing is carried out using version 1.8.2 of the JWST pipeline \citep{JWST_pipeline} and v0995 of the Calibration Reference Data System (CRDS), which were most recent in the first half of 2023. (2) Wisps and artefacts from F150W and F200W are subtracted using a template set in between stages 1 and 2 of the pipeline. (3) 1/f noise correction, derived by Chris Willott.\footnote{\url{https://github.com/chriswillott/jwst}}, is applied after stage 2 of the pipeline. (4) A 2-dimensional sky subtraction is run on each of the individual calibrated frames before stacking in stage 3 of the pipeline. (5) After stage 3 of the pipeline, the final F444W image was matched to a GAIA-derived WCS using available GAIA stars in the NIRCam imaging, and all other images in the other bands were then aligned with the new F444W image to ensure consistent source positions. The processed images have a final resolution of 0.03 arcsec/pixel.

\subsection{Initial high redshift catalog construction} \label{sub:ini_cat}

Source detection and measurement from the processed science images was carried out using SExtractor \cite{sextractor}, with the key configuration parameters taken from Table 1 in \cite{2023MNRAS.518.4755A}. We used the F444W band for initial source detection.  Using this fluxes of the detected sources were then measured in each band using 0.32 arcsec diameter apertures, with corrections applied derived from Point Spread Function (PSF) models taken from \cite{WebbPSF}. Detection depths were calculated individually for each source by placing 0.32 arcsec diameter apertures in empty spaces of the image, then picking 200 nearest apertures for each source and calculating the total background RMS across all of them. The 5$\sigma$ detection depth was estimated as 5 times the background RMS in magnitudes. A summary of average 5$\sigma$ depths is provided in \autoref{tab:5sigma}.  This is further described in \cite{UV_LF_methods} and Conselice et al. (2023, in prep).

Initial source redshifts were constrained by photometrically fitting the SExtractor catalog sources with both the LePhare and EAZY \cite{EAZY} codes. All detected sources were run through these SED fitting tools in order to provide a photometric redshift estimate.   Both EAZY and Le Phare were run with a minimum 5\% flux error to account for uncertainties in the calibration of the NIRCam detector. 

The LePhare code was run using the BC03 stellar population synthesis (SPS) template set \citep{Bruzual2003} with both exponentially decaying and constant star formation histories (SFHs). We include templates with 10 characteristic timescales between $0.1<\tau<30~\mathrm{Gyr}$, and 57 different ages between 0 and~13 Gyr, with fixed metallicities $Z=\{0.2,~1.0\}~\mathrm{Z}_{\odot}$. The redshift range is allowed to vary between $0<z<25$, and dust extinction is varied from 0 < $E(B-V)<3.5$ to account for potential dusty lower-z contaminants \citep[e.g.][]{Naidu2022b,Zavala2023}. Attenuation from the inter-galactic medium (IGM) follows the treatment derived in \citep{Madau1995} and nebular line emission is modelled internally by Le Phare. 

EAZY was run with the 12 default Flexible Stellar Population Synthesis {\tt fsps} templates { \tt (tweak\_fsps\_QSF\_v12\_v3)}, which model a range of stellar ages, dust extinction and metallicities \citep{conroy2010fsps}, along with 6 additional templates from \citet{Larson2022}. These templates were designed to better reproduce the blue colors and $\beta$ slopes of high-z galaxies. Some high-z galaxies have been shown to have high equivalent width (EW) emission lines, which are more accurately modelled by the FSPS templates, which can boost photometric measurements by as much as a magnitude. This EAZY template set will be referred to as FSPS+Larson hereafter.

The selection criteria for constructing the robust high-redshift galaxy catalog can be summarized as follows, although see Conselice et al. (2023, in prep) and \citep[][]{UV_LF_methods}:

\begin{itemize}
    \item The candidate object must have a higher than 5$\sigma$ detection in the first two bands redwards of the fitted Lyman break and < 3$\sigma$ detections in bands containing the break.
    \item The integrated probability density function (PDF) within 10 \% of the best fit redshift must contain at least 60\% of the probability.
    \item Less than 10 \% of the PDF must lie in the $z < 5$ range; secondary peak solutions, if present, must have a maximum lower than 50 \% of the primary peak to avoid Lyman - Balmer break confusion.
    \item Candidate must have $\chi^2_{R} < 6$ to be considered `good' or $\chi^2_{R} < 3$ for a 'robust' classification.
    \item The PDF criteria must be satisfied by both codes and their photometric redshifts have to be consistent within 3$\sigma$.
\end{itemize}

The above procedure is discussed in depth by \cite{UV_LF_methods}.

A total of 214 high-redshift ($6.5 < z < 12$) sources were identified using our criteria and were further analyzed for the presence of an AGN component. It should be noted here that the lowest redshift available in CEERS and NGDEEP surveys was 8.5 instead as these fields use F115W as the bluest available band and we did not incorporate the Hubble Space Telescope (HST) imaging into our selection.

\begin{table*}
    \centering
    \begin{tabular}{cccccccc}
    \hline
        Band & CEERS & GLASS & NGDEEP& SMACS-0723 & MACS 0416 &  El Gordo & NEP \\
    \hline
        F090W & -- & 28.90 & --      &28.80& 28.70 & 28.30 & 28.50 \\
        F115W& 28.75 & 28.95  & 29.65 &--&28.65 & 28.30& 28.50  \\
        F150W& 28.60 & 28.50 &29.50  &28.75&28.50 & 28.15 & 28.60\\
        F200W& 28.80 & 28.80 &29.50  &28.90&28.65 & 28.40 & 28.70 \\
        F277W& 28.95 & 29.15 &29.80  &29.25&28.95 &28.80 & 29.00\\
        F356W& 29.05 & 29.30 &29.75 &29.40&29.10 &28.85 & 29.10\\
        F410M& 28.35 & --  &--       &--&28.65 &28.35 & 28.45\\
        F444W& 28.60 & 29.50& 29.55 &29.15&28.90 &28.70 & 28.75  \\
    \hline
    \end{tabular}
    \caption{A summary of average 5$\sigma$ detection depths of each NIRCam filter for the surveys examined in this work. It should be noted that due to bright foreground objects and edge noise effects these values varied by up to 1 mag across the images and thus were examined individually for each candidate object. Bands that did not have data for a particular field are marked with '--'.}
    \label{tab:5sigma}
\end{table*}

\section{Candidate AGN selection}\label{sec:cand}

In the following section we describe how our AGN candidates were found using our methods based on the full EPOCHS sample described above.  This involves several step including the initial discovery of the objects through a series of photometric redshift codes and tests see \cite[][]{UV_LF_methods} and the previous section for further details. We then carry out a further analysis examining the likelihood that these systems are dominated by emission from black holes to construct our final sample.

\subsection{SED fitting} \label{sub:SED}

This work seeks to identify robust candidate AGN. The easiest ones to identify using imaging are those of the unobscured, Type 1, variety, where the immediate surroundings of the black hole are capable of outshining its host galaxy. Our methods are designed to identify these particular AGN candidates and subsequently are not expected to produce a complete sample of the AGN population in the data covered by the EPOCHS sample.

We begin our selection by refitting our sources from the robust galaxy catalog using EAZY with a set of SED templates for direct collapse black hole (DCBH) hosts from \cite{DCBH_templates} added on top of the FSPS+Larson set (see section \ref{sub:ini_cat}). These templates are tuned for unobscured, intermediate mass ($10^5 $ - $10^6$~M$_{\odot}$) active black holes, which may reasonably be expected to make up a significant fraction of high redshift AGN.  This gives us an AGN+star formation set of templates from which we can find galaxies that match various combinations of these templates. The continuum shape of these SEDs is characterized chiefly by their UV power-law slopes ($\alpha$), and the so called Big-Bump temperatures ($T_{bb}$). We adopt the full range of values for both, with $\alpha = -1.2, -1.6, -2.0$ and $T_{bb} = 5\times10^4, 1\times10^5, 2\times10^5 K$. We fix the ionization parameter $\log{U}$ to -0.5 and metallicity $Z$ to 0.014 as reasonable choices for pristine, high redshift environments \citep{Sarmento2017}. These parameters are otherwise difficult to constrain using SED fitting. Thus, the additional set of SEDs consists of 9 templates with the aforementioned parameters and will be referred to as the 'Nakajima' set hereafter.

 After re-fitting, we derive the weighting for each template in the fit via the following equation:
 \begin{equation}
 \label{eq:weights}
     W_i = a_i\left(\sum_ja_j\right)^{-1},
 \end{equation}
 where $a_i$ is the linear coefficient of the i-th template as defined in \cite{EAZY}. From these, we define the $W_{AGN}$ parameter as $W_{AGN} = \sum W_i$ for all AGN templates in the set. This parameter thus serves as an indication of the relative weight of AGN versus non AGN templates in the best fit for each source.
 
 Sources were then selected according to the following criteria:
 \begin{itemize}
     \item $\chi^2_{R} < 3$, consistent with the 'robust' classification from section~\ref{sub:ini_cat}.
     \item Nakajima templates having $W_{AGN} > 0.5$ in the fit, ensuring that a candidate is mostly fitted by an AGN model.
     \item The new $\chi^2_{R}$ value is lower by at least 0.5 than the one given by FSPS+larson set to ensure that the fit improvement provided by adding the AGN models is not the result of an expanded parameter space.
     \item Redshift given by the Nakajima templates consistent with other redshift estimates as the location of the Lyman break should be insensitive to the nature of emission. 
 \end{itemize}
 
\begin{figure}
    \centering
    \includegraphics[width=\columnwidth]{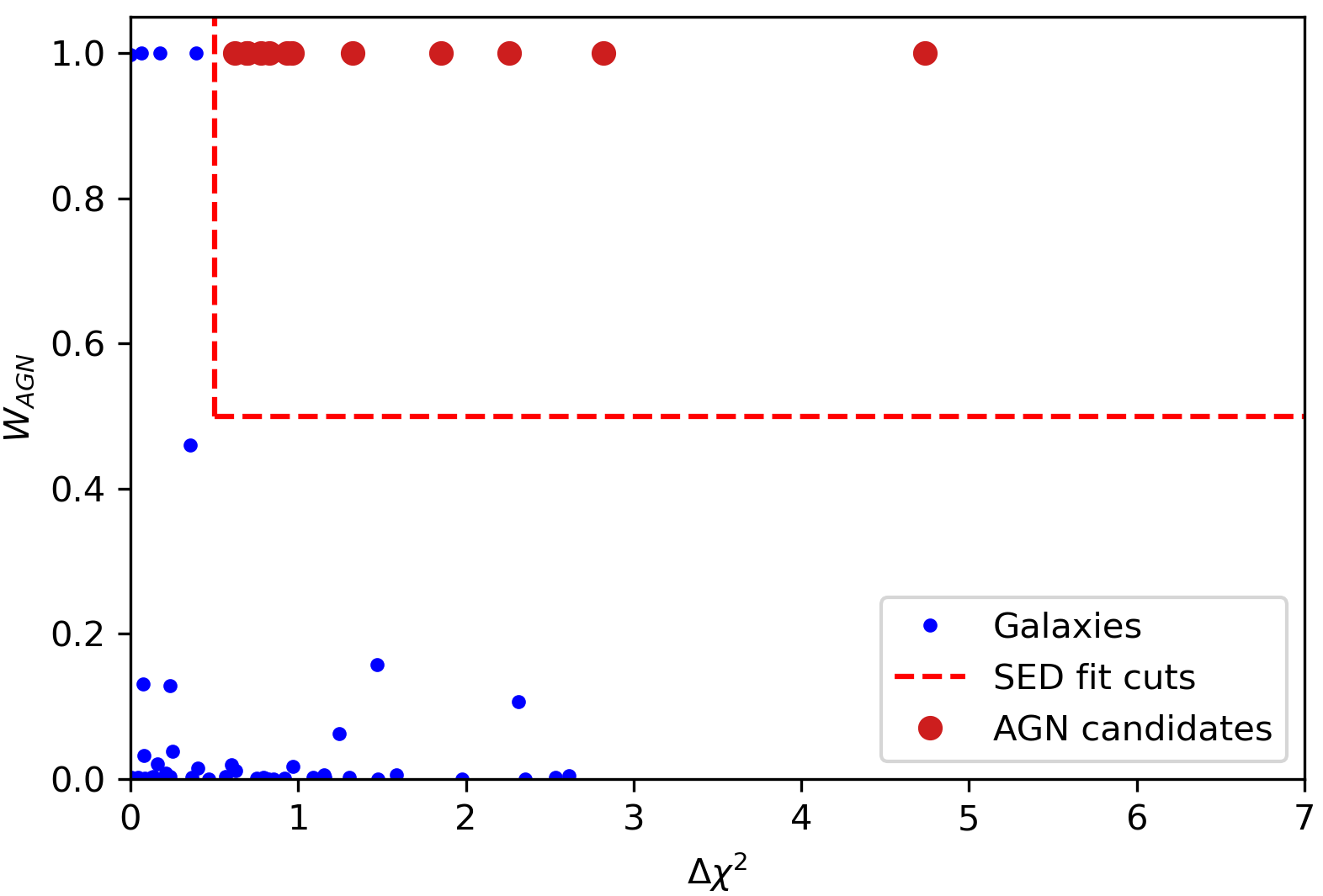}
    \caption{Graph illustrating the separation of our pre-selected candidates from the rest of the sample in the parameter space. The weights of the Nakajima templates ($W_{AGN}$) are plotted on the y-axis, the x-axis shows the difference between the $\chi^2_R$ values given by the Nakajima and FSPS+Larson sets ($\Delta\chi^2$). Only sources with $\chi^2_R < 3$ given by the Nakajima templates are plotted.}
    \label{fig:sed_cuts}
\end{figure}

The above procedure resulted in 12 sources being selected from the initial sample. This selection is illustrated by \autoref{fig:sed_cuts}. The AGN candidates are strongly separated from the rest of the sample along the $W_{AGN}$ axis, with most sources concentrated either at 1 or 0. This is likely because EAZY prefers single template models instead of mixed templates. Therefore, this parameter does not  necessarily correspond to a physical AGN fraction, but it remains useful for further selection of strong candidates.

Afterwards, as part of our refined method for finding AGN, we then use CIGALE \citep[][]{Cigale} to narrow our selection. Before fitting our SEDs to these templates we increase the SExtractor measured flux errors such that they have a floor, or lower limit values, which represent a 5\% error.  We then set the 3$\sigma$ upper limits with 1$\sigma$ errors in bands which contain fluxes consistent with a 3$\sigma$ background fluctuation. Upper limits of 1$\sigma$ with 1$\sigma$ error were set in bands containing negative flux measurements. 

We modeled the AGN component of the CIGALE templates using the SKIRTOR continuum emission models \citep{Skirtor_mods} with a varying rest-frame UV slope ($\alpha$) as the key shape parameter.  This was chosen to largely overlap with the $\alpha$ values from the Nakajima set. The allowed viewing angles were 30 and 70 degrees in order to consider both obscured and unobscured emissions. The stellar emission was modeled using an initial mass function from \cite{Salpeter-IMF} together with the BC03 templates, and a delayed exponential star formation history, with stellar ages ranging from 5000 to 100 Myr, and $0.5<\tau<2~\mathrm{Gyr}$. This range is narrower than the one used with LePhare due to the need to simplify the parameter space to allow for more AGN models and the high redshift nature of the fitted galaxies already being confirmed by the previous selection steps. 

The stellar and gas metallicities in our fit were sampled from the range $Z=\{0.2,~1.0\}~\mathrm{Z}_{\odot}$, consistent with what used within the LePhare fitting. The nebular emission was modeled with the ionization parameter using values of -1.0, -1.5 and -2.0. The dust extinction for the stellar component was modeled using the Calzetti dust attenuation law, assuming $0 < E(B-V) < 0.9$. AGN polar dust extinction was assumed to follow the SMC curve, taken from \cite{SMC}, with extinction values ranging from $0 < E(B-V) < 0.8$. The values not listed were kept to CIGALE defaults.

The relative performance of AGN versus non-AGN models was established by running CIGALE with two groups of SED templates, the first one with $f_{AGN} = 0$, while the second ranges from  $0.1 < f_{AGN} < 1$. This $f_{AGN}$ parameter quantifies the ratio of observed infrared luminosity of the AGN component to the total observed infrared luminosity of the source. The average performance of the two template sets was quantified using a parameter $P_{AGN}$:
\begin{equation} \label{eq:Pagn}
    P_{AGN} = \frac{N_{AGN}(\chi^2_R<\chi^2_{lim})}{N_{GAL}(\chi^2_R < \chi^2_{lim})}\times\frac{N_{GAL}}{N_{AGN}},
\end{equation}
where $N_{AGN}(\chi^2_R<\chi^2_{lim})$ is the number of AGN models giving $\chi^2_R<\chi^2_{lim}$, and $N_{GAL}(\chi^2_R<\chi^2_{lim})$ is the number of galaxy models satisfying the same criterion, $N_{AGN}$ and $N_{GAL}$ are the total number of AGN and galaxy templates fitted.  We use this ratio to normalise the number of models as otherwise one type would dominate over the other.   In cases where $N_{GAL}(\chi^2_R<\chi^2_{lim}) = 0$ and $N_{AGN}(\chi^2_R<\chi^2_{lim}) \neq 0$, $P_{AGN}$ was set $=99$, if no models gave $\chi^2_{R} < \chi^2_{lim}$, $P_{AGN}$ was set to $=0$. Sources with $P_{AGN} > 1$ were selected for furhter morphological 
and structural analysis. 

The value of $\chi^2_{lim}$ was fixed by using different values $\chi^2_{lim}$ for \autoref{eq:Pagn} to classify a sample of known AGN and likely non-AGN sources, minimizing the number of misclassifications.  For the known AGN sample we use three objects in total - the two spectroscopic AGN from \cite{hidden_monsters}, CEERS 1670 and CEERS 3210, and one from \cite{CEERS_z8_agn}, CEERS 1019, as these low mass sources are more likely to be representative of the ultra high redshift AGN population. None of these sources end up in our robust catalog due to them having >3$\sigma$ detections in the F115W band where the Lyman-break is located, causing them to fail one of the robust redshift criteria in section \ref{sub:ini_cat}.  In fact objects such as these objects require HST data for reliable classification to ensure that there is a Lyman-break and not a Balmer-break within the 'drop' filter. However, we use our measured photometry from the original SExtractor catalog for fitting this object. We also note that CEERS 3210 is obscured, while CEERS 1670 is a classic, more evolved, Type 1 AGN, thus the Nakajima templates, calibrated for AGN hosted by pristine early environments, do not reproduce their photometry well. The CEERS 1019 object, while having a relatively flat continuum has a strong OIII line visible in the F444W band, which is not captured well by any SED templates used in our fitting. This results in a fit which does not imply an AGN as the templates we use do not have line emission this strong.  This reveals that even stronger line emission from AGN should be implemented in future AGN template models. 
 This source also has a continuum strongly influenced by stellar emission, see discussion in \cite{CEERS_z8_agn}. 

\begin{figure}
    \centering
    \includegraphics[width=8.5cm]{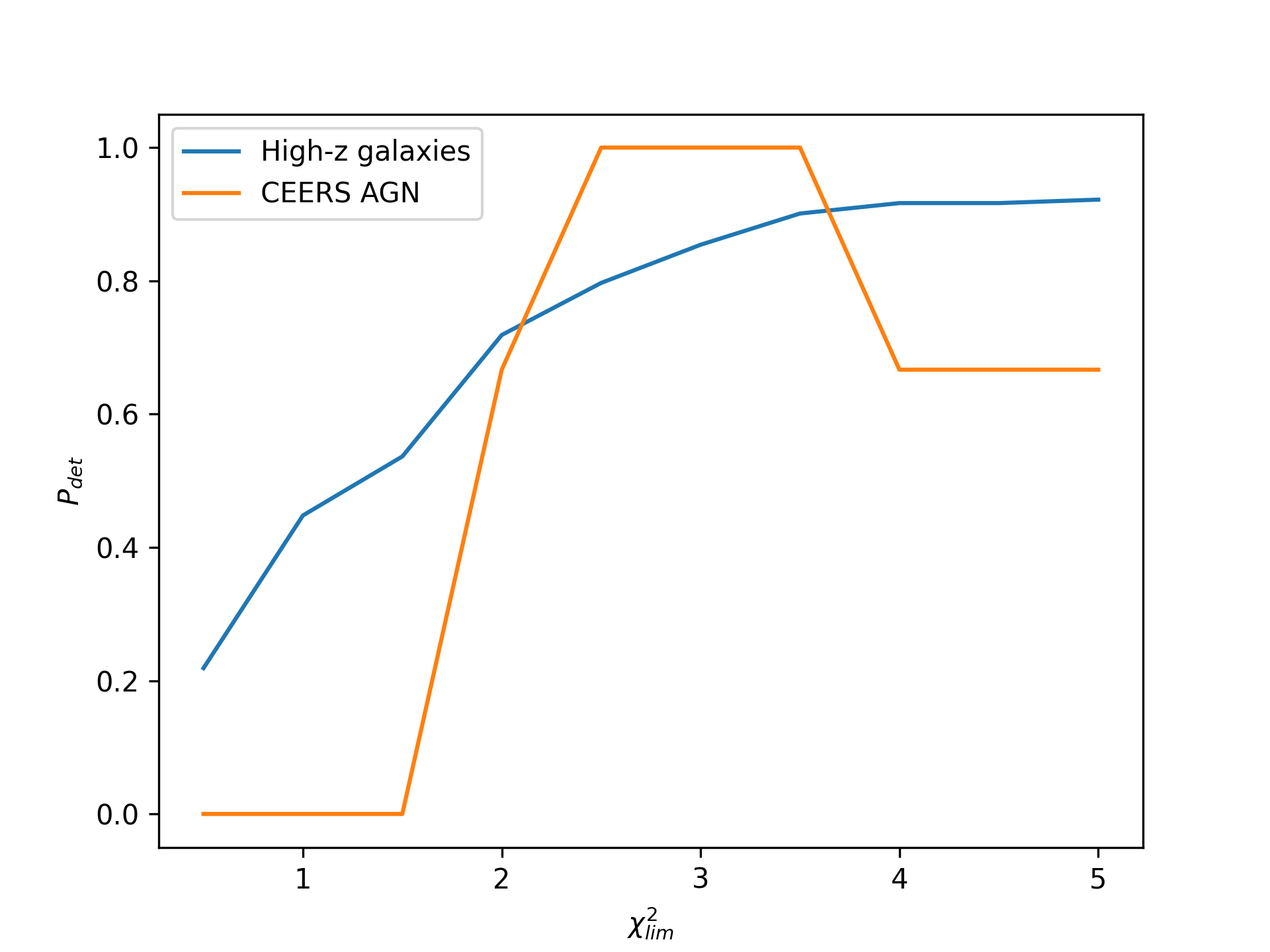}
    \caption{A graph showing the dependence on the probability of classifying a source as an AGN ($P_{det}$), and the value of $\chi^2_{lim}$ for the samples of CEERS AGN (orange line) and High-z galaxies (blue line). It can be seen that $\chi^2_{lim} = 2.5$ gives the maximum chance of correctly classifying an AGN while minimizing the chance of incorrect classification. The probability of incorrect classification remains relatively high, however, at 80\%.}
    \label{fig:chi_lim}
\end{figure}

The non-AGN high redshift galaxy sample was taken from the original robust galaxy catalog by removing all galaxies that satisfied the EAZY selection criterion and were not classified by us as AGN using our methods.  
 We run these galaxies through our procedure and the results of this  are shown in \autoref{fig:chi_lim} as the blue line.  We find that some of these galaxies do have a high AGN fraction, and thus it remains possible, if not likely, that some of these systems are in fact AGN. However, using our methods we are more certain to find a pure selection of AGN as also shown by the orange line. 

As can be seen in \autoref{fig:chi_lim}, $\chi^2_{lim} = 2.5$ gives the highest probability of correctly classifying an AGN, however, 80\% of the remaining galaxy sample has $P_{AGN} > 1$, and while some of these sources may harbour obscured AGN akin to CEERS 3210, such a high AGN fraction is unlikely as the fraction of dust reddened AGN at $4<z<7$ was estimated to be about 10\% by \cite{Harikane2023_agn}. Therefore, this method can only be used to exclude dubious sources as its purity is too low for standalone use. Nevertheless a further two sources are excluded from the sample by this method.   

This is our main method for finding AGN. It is however important to note that this is not a unique method, and other methods using photometry and SED fitting can be developed.  However our method does provide a way for finding a sample with a high probability of being AGN. It is also important to note that the template set with which we constrain most of our sources was produced using unobscured AGN models, thus our search is inherently biased towards Type 1 AGN in unevolved low metallicity environments.

\subsection{Structural Measures}

\begin{figure}
    \centering
    \includegraphics[width=\columnwidth]{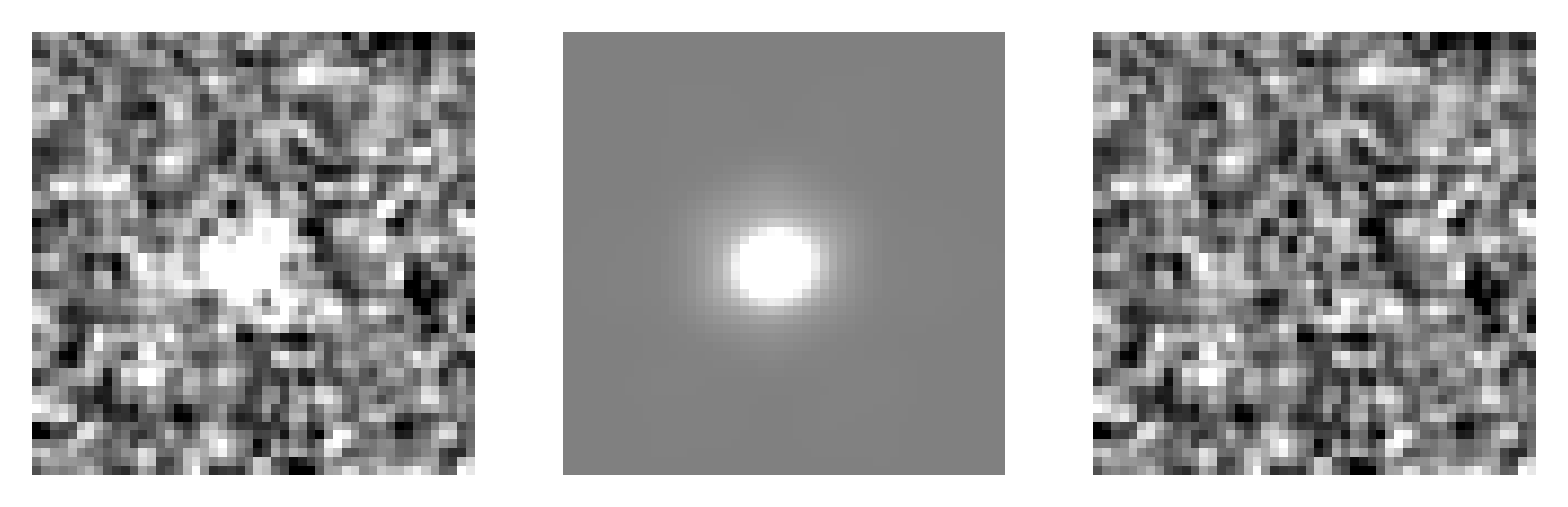}
    \small (a) NEP-z7l
    \includegraphics[width=\columnwidth]{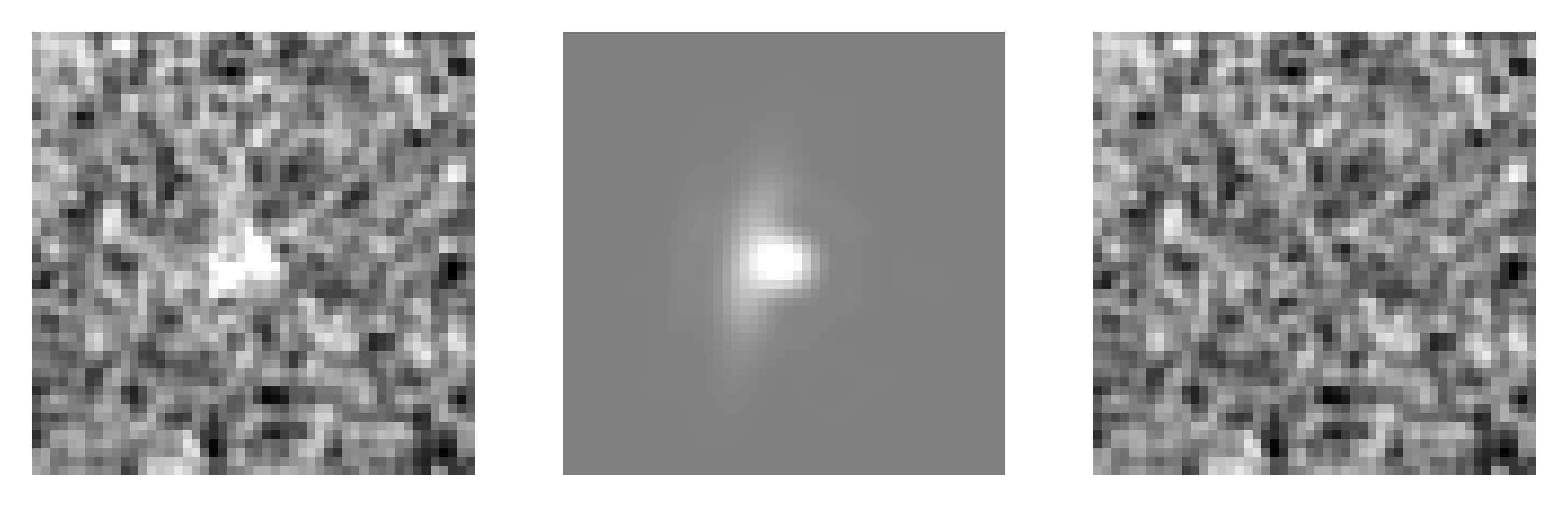}
    \small (b) S0723-z11c
    \includegraphics[width=\columnwidth]{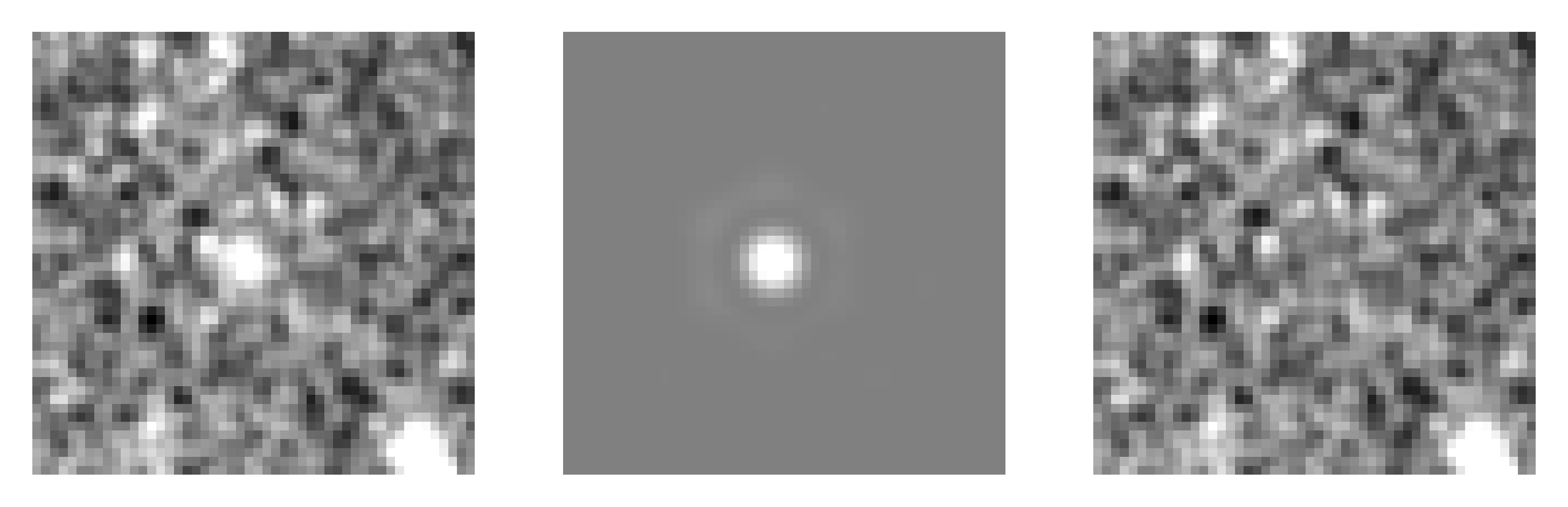}
    \small (c) CEERS-z9d
    \caption{Top: An example morphological fit of NEP-z7l, which is best fit by a Sersic profile, which is compact enough to be classified as an AGN, with $R_e < $ F444W PSF FWHM.
    Middle: S0723-z11c, best fit by a combined Sersic and PSF model whose centres do not coincide.  This system is very unusual in structure and we discuss this object later in this paper.
    Bottom: CEERS-z9d, best fit by a PSF. 
    The left panels show the data images, the centre panels show the PSF models, and the right panels show the residual images, created by subtracting the model image from the data image. The residual images show these models are good fits, as there are no bright areas of remaining light unaccounted for, or any dark areas over-fit by the model. All cutouts are 1.5'' x 1.5''.}
    \label{fig:morph}
\end{figure}

In order to improve the purity of our sample, we apply morphological cuts to search for sources containing a point source. To do this, we use \galfit, a two-dimensional fitting algorithm that uses a Levenberg-Marquardt algorithm to find the optimum solution to a fit \citep{Galfit1, Galfit2}.  We select sources which are best fit by a point spread function (PSF), a combination of an extended Sersic profile and a PSF, or a single Sersic profile with half light radius less than the FWHM of the PSF used in the fitting process. We define the best-fitting model as the model with the lowest $\chi_v^2$, which is defined by \galfit\ as 
\begin{equation}
\chi_\nu^2=\frac{1}{N_{\mathrm{DOF}}} \sum_{x=1}^{n_x} \sum_{y=1}^{n_y} \frac{\left(f_{\mathrm{data}}(x, y)-f_{\mathrm{model}}(x, y)\right)^2}{\sigma(x, y)^2}.
\end{equation}
Within our fitting we use the SExtractor parameters for each object as the initial guesses, and run \galfit\ three times for each object, modelling the source as an extended Sersic profile, a Sersic profile containing a PSF, and as just a pure PSF. For all fits, the ERR extension of the image, which is a measure of the noise of the image, is used as the input sigma image. Sources containing an AGN cannot be modelled accurately by a single Sersic fit, as the AGN appears as a distinct point source. However, determining the structures of sources with angular sizes similar to the PSF of the telescope is difficult and results should be interpreted with some caution \citep{dewsnap2023}. As a result of this, we also select sources where the determined half light radius is less than the FWHM of the PSF. All object cutouts are from the F444W NIRCam image, where the PSF for this band has a FWHM of 4.83 pixels on our pixel scale, therefore sources with $R_e < $ 4.83 pixels are selected as being a point source object. We use the F444W band for our fitting process, as this is closest to the rest-frame optical for each source, and keep this consistent throughout in order to model each source using the same parameters and constraints. Due to the faint magnitudes of these sources, we fix the Sersic index to a value of $n = 1$. Where multiple sources are fit simultaneously, the image positions of all objects are constrained to within $\pm$ 3 pixels, to ensure the correct sources are fit. We also visually inspect fits and residual images as a final quality check. An example of each fit is shown in in \autoref{fig:morph}.

The final structural analysis result in two of our objects being classified as a PSF, five objects being classified as a Sersic profile with $R_e < $ the FWHM of the PSF, and four objects classified as a combined model of an extended Sersic profile containing a PSF. The remaining two objects are not morphologically classified as a pure AGN due to their radius being larger than that of the PSF.   These could in fact still be AGN, but we are interested here in systems where the AGN dominates the light of the source.  The classification of each object is given in \autoref{tab:morph}, and properties, including radii for those fit as Sersic profiles, are given in \autoref{tab:sources}. 

We further check our results in the F277W band, where most of our sources have higher S/N ratios, and find that our classifications do not change. In particular, we check the sizes of our sources best fit by a single Sersic profile, and find that in general, we recover them as having sizes smaller than the FWHM in F277W. We find one source as having $R_{e, F277W} \sim 1.04\times\textrm{FWHM}$, which could occur due to noise, or faint extended emission better detected in this band, and as such, we still classify this as a compact Sersic profile, small enough to be a PSF. The only source we do not recover in this way, is CEERS 1019, however this source has a very complex morphology, which we analyse further. 

We find that the source discovered in \citet{CEERS_z8_agn} is classified differently in our methods than in the discovery paper, where the object is found to be three components, with the central component best fit by a combination of a PSF and Sersic profile. Our combined fit of a Sersic profile and PSF has a marginally higher $\chi_{\nu}^2$, therefore we select this object as a compact Sersic profile, small enough to be classified as a PSF. This source has a complex morphology due to likelihood of it being a merger, and thus we replicate the fitting process completed in the discovery paper, and model the source as a three component model, with two components fit by Sersic profiles, and a central PSF component. We find that this has a lower $\chi^2_{\nu}$ than our model fits, confirming our original findings that this object is compact enough to contain a point source. Our final classification information for the 12 sources selected from Nakajima templates is given in \autoref{tab:class}.

\begin{table}
    \centering
    \setlength{\tabcolsep}{3.7pt}
    \begin{tabular}{cccc}
    \hline
    PSF & Sersic & Combined & Not Selected \\
    \hline 
    NEP-z9b & NEP-z7l & S0723-z11c & NGD-z8f \\ 
    CEERS-z9d & NEP-z8ff & CEERS-z6c & NEP-z7m \\
    - & CEERS-z8d & NEP-z6c & -\\
    - & CEERS 1019 & NGD-z8e & -\\
    -  & CEERS-z8g & - & - \\ \hline
    \end{tabular}
    \caption{A table summarizing the morphological classification of each object within our sample. To be classified as a Sersic fit, an object must have $R_e$ < 4.83 pixels, the FWHM of the F444W PSF re-sampled to our pixel scale. Both objects that are not selected are rejected due to their best fit model having $R_e$ larger than the PSF FWHM.}
    \label{tab:morph}
\end{table}

\begin{table*}
    \begin{tabular}{ccccccccc}
    \hline
        ID & $\chi^2_{Naka}$ & $\Delta \chi^2$ & $P_{AGN}$ & $\chi^2_{GALFIT}$ & $f_{PSF}$ & $R_e$ (px) &z &Selected? \\
    \hline
         NEP-z7l & 0.96 & 2.8 & 1.3 & 5.093 & 0  & 2.87 $\pm$ 0.66 &$8.2^{+0.5}_{-0.3}$& Yes \\ 
         CEERS-z8i& 1.91 & 0.70 & 1.6 & 1.178 & 0.19 $\pm$ 0.02 & 5.58 $\pm$ 0.67 &$8.8^{+0.3}_{-0.3}$& Yes \\ 
          NEP-z9b & 1.04 & 1.9 & 1.4 & 10.61 & 1 & $< 4.83$ &$9.5^{+0.7}_{-0.7}$& Yes \\ 
          NEP-z8ff& 0.55 & 2.3 & 1.4 & 0.801 & 0 & 2.34 $\pm$ 0.34 &$8.7^{+0.2}_{-0.3}$& Yes \\ 
          NEP-z7m& 1.71 & 4.7 & 1.2 & 1.18 & 0 & 5.15 $\pm$ 0.44 &$7.3^{+0.1}_{-0.1}$& No \\ 
          NEP-z6c & 1.83 & 0.62 & 1.2 & 0.887 & 0.17 $\pm$ 0.02 & 8.04 $\pm$ 1.14 &$6.5^{+0.2}_{-0.2}$& Yes \\ 
          CEERS-z8d&1.74 & 0.83 & 1.4 & 0.592 & 0 & 4.64 $\pm$ 0.42 &$8.8^{+0.2}_{-0.2}$& Yes \\ 
          S0723-z11c& 0.77 & 0.93 & 1.3 & 0.939 & 0.36 $\pm$ 0.03 & 5.69 $\pm$ 1.1  &$11.9^{+0.3}_{-0.4}$& Yes \\ 
          NGD-z8e & 2.75 & 1.3 & 0 & 2.273 & 0.34 $\pm$ 0.04 & 5.74 $\pm$ 1.85 &$8.9^{+0.4}_{-0.3}$& No \\ 
          CEERS-z8g & 0.14 & 0.69 & 1.3 & 0.977 & 0 &4.14 $\pm$ 0.62 &$8.8^{+0.3}_{-0.3}$& Yes \\ 
          NGD-z8f & 1.39 & 0.78 & 0.9 & 2.594 & 0 & 8.09 $\pm$ 2.8 &$8.9^{+0.3}_{-0.3}$& No\\ 
          CEERS-z9d&1.38 & 0.96 & 1.5 & 0.738 & 1 & $< 4.83$ &$9.0^{+0.3}_{-0.3}$& Yes \\ 
          \hline
    \end{tabular}
    \caption{Classification information for the 12 sources initially selected from fitting the Nakajima template set. The first column lists our catalog ID, the following columns present the parameters used to classify each source as an AGN: $\chi^2_{Naka}$ - the value of $\chi^2$ given by the Nakajima AGN template set, $\Delta \chi^2$ - the difference between best fit $\chi^2$ of the Nakajima and FSPS+Larson template sets, $P_{AGN}$ - the parameter defined in \autoref{eq:Pagn}, $\chi^2_{GALFIT}$ - best fit $\chi^2$ given by {\tt{GALFIT}} and $f_{PSF}$ - fraction of total flux contained in the compact component, $f_{PSF} = 1$ implies the best fit morphology is a PSF, $f_{PSF} = 0$ implies a Sersic profile best fit and $f_{PSF} < 1$ implies the best fit morphology is a Sersic profile containing a PSF. Column labelled $R_e$ contains the the half-light radii in pixels for sources best fit by a Sersic profile, errors on these are standard {\tt{GALFIT}} errors. Where a source is best fit by a combined model, this is taken from the half light radius of the extended Sersic profile, and for those best fit by a PSF, the radius is the upper limit given by the PSF FWHM. Second to last column contains the photometric redshifts given by LePhare. The final column indicates if the source was selected as an AGN.}
    \label{tab:class}
\end{table*}

\section{AGN Source properties}\label{sect:src}

Using our selection procedure we identify a total of nine robust candidate sources out of a sample of 214. We also include the CEERS 1019 source from \cite{CEERS_z8_agn} for the sake of comparison with our candidates, for a total sample of ten high redshift sources. Thus we estimate an AGN fraction at $ 6.5 < z < 12$ of $5\pm1$\%, assuming a Poisson counting error. Due to our investigation focusing on purity rather than completeness as well as being strongly biased towards Type 1 AGN, this value should be viewed as very much a lower bound estimate. This is still consistent with the 1 - 10\% observable AGN fractions derived from the FLARES simulation results by Kuusisto et al 2023 (in prep) and matches well with \cite{Harikane2023_agn} finding of $\sim$5\% of galaxies at $4<z<7$ hosting low luminosity Type 1 AGN, potentially indicating that the AGN fraction does not evolve much during this epoch.

The $f_{AGN}$ values for our sources were estimated by rerunning CIGALE with the same parameters as in section \ref{sub:SED}, except the $f_{AGN}$ parameter was varied over the full range of 0 to 1 in steps of 0.1. Physical values of $R_e$ were measured by noting that the pixel scale of the images was 0.03 arcsec, and using angular diameter distances calculated at best-fit redshifts, with both {\tt{GALFIT}} and redshift errors contributing to the final uncertainties. The values of $T_{bb}$ and $\alpha$ were taken from the best fitting Nakajima templates, the model grid for these being too sparse to estimate meaningful uncertainties. We also measure the rest-frame absolute UV magnitudes $M_{UV}$ by redshifting the best-fit SED to $z = 0$ and convolving it with a top-hat between 1450 and 1550 \r{A} in wavelength space, with the uncertainties provided by redshift errors. Photometric redshifts and their errors were taken from LePhare results. All physical properties measured for our candidate sources are presented in \autoref{tab:sources}.

\begin{figure*}
\centering
\subfloat{\includegraphics[width=\columnwidth]{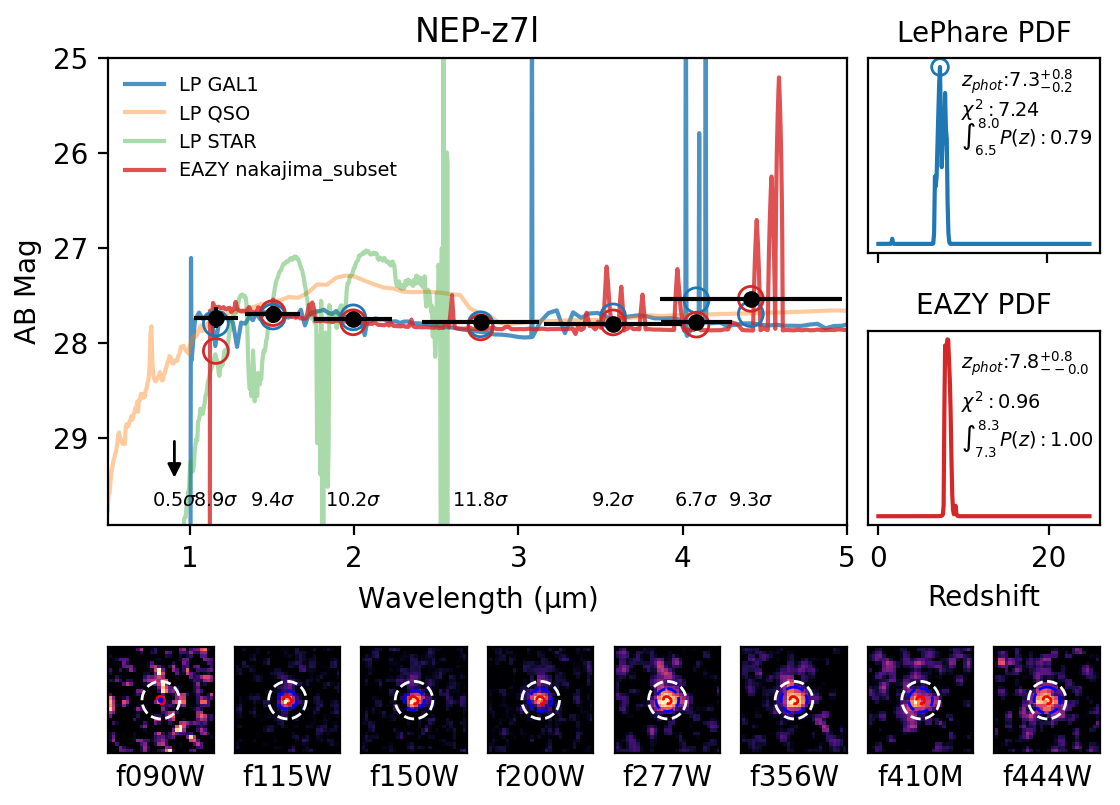}} \hfill
\subfloat{\includegraphics[width=\columnwidth]{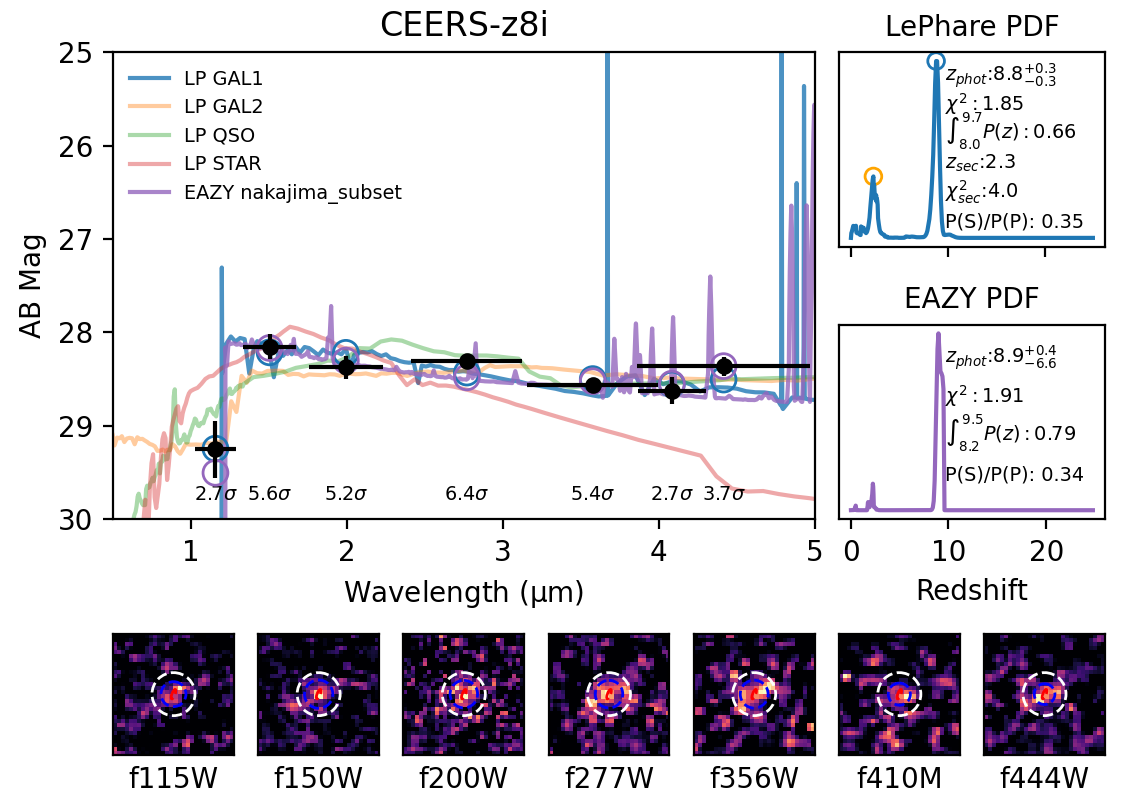}}\hfill
\subfloat{\includegraphics[width=\columnwidth]{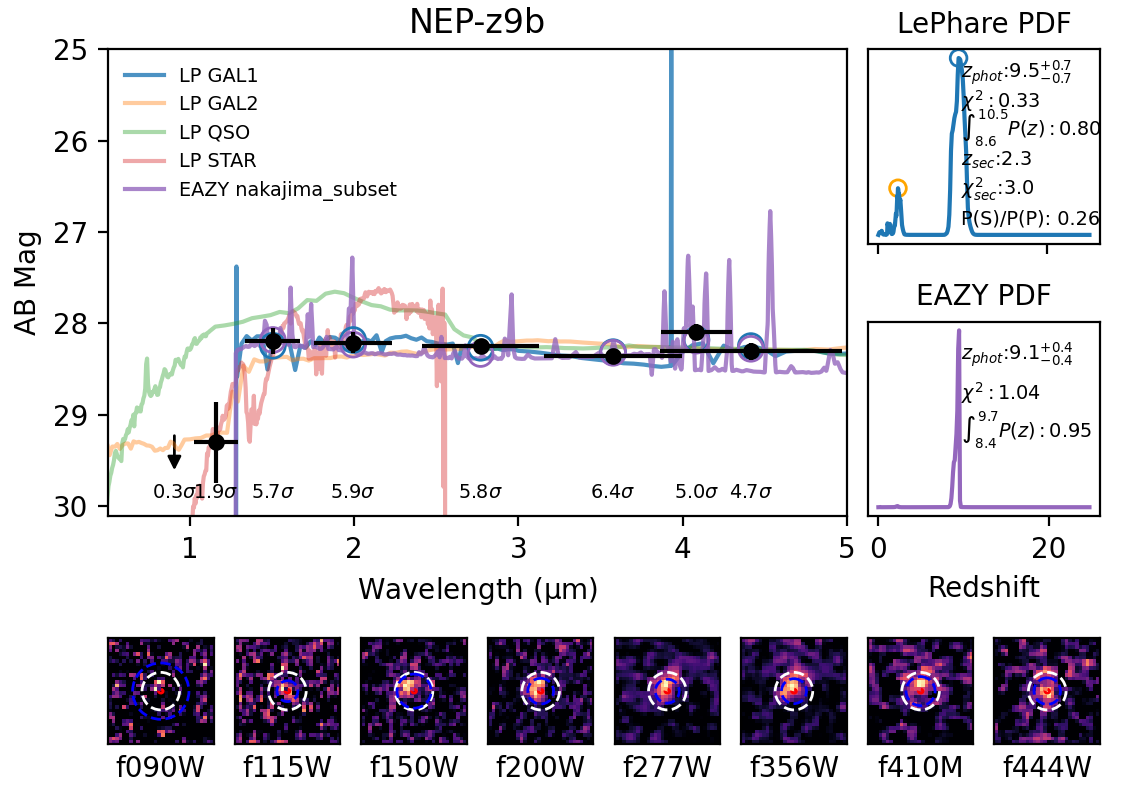}}\hfill
\subfloat{\includegraphics[width=\columnwidth]{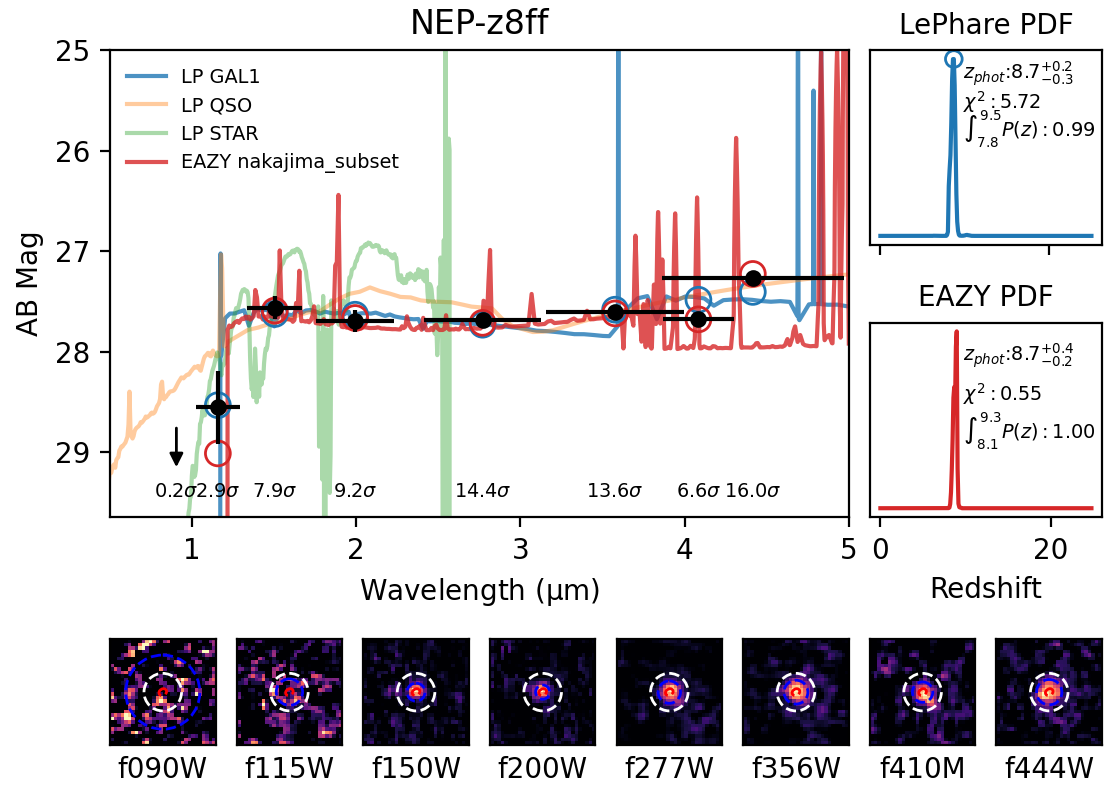}}\hfill
\caption{Best fit SEDs for the first 4 of our 9 candidates. The best fitting LePhare galaxy SED is shown in blue, the best fit Nakajima SED in red for sources without a secondary solution identified by LePhare and purple for the others. PDFs obtained from both fits are shown to the right. The image cutouts of the source in all measurement bands are presented below the plots. The white circles show the position of the 0.32 arcsec apertures, the blue ones are the half light radii measured by SExtractor. The top left source, NEP-z7l, stands out as particularly well fitted by an AGN SEDs, with $\Delta\chi^2 > 2$ when compared to both FSPS+Larson and LePhare templates, with broad hydrogen line emissions explaining the observed excess in the F444W band. This source is also best-fit by a PSF-like morphology. The general trend among our candidates is that broad AGN emission lines tend to better explain the slight excess fluxes in the red bands, leading to better $\chi^2$ values. Image cutouts reveal most sources to either have a compact point-like nature or exhibit signs of a bimodal structure, indicating potential mergers or disturbed morphology.}
\label{fig:SED_1}
\end{figure*}
\begin{figure*}
\centering
\subfloat{\includegraphics[width=\columnwidth]{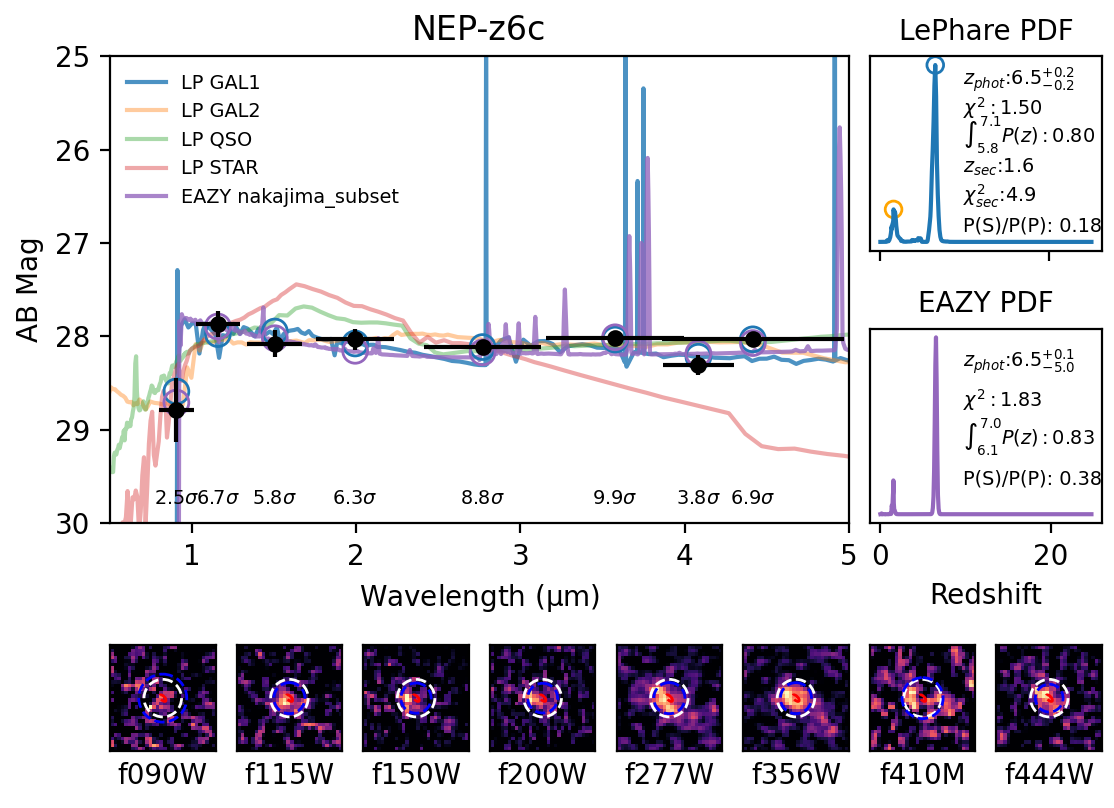}}\hfill
\subfloat{\includegraphics[width=\columnwidth]{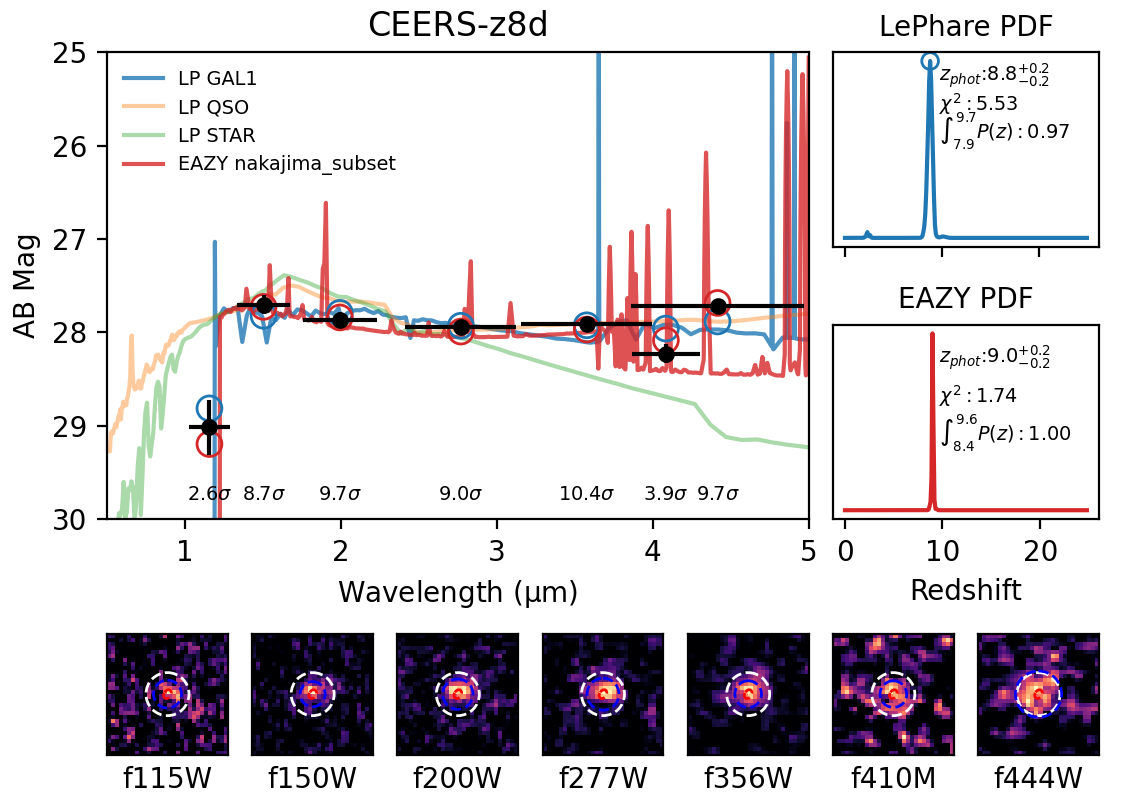}} \hfill
\subfloat{\includegraphics[width=\columnwidth]{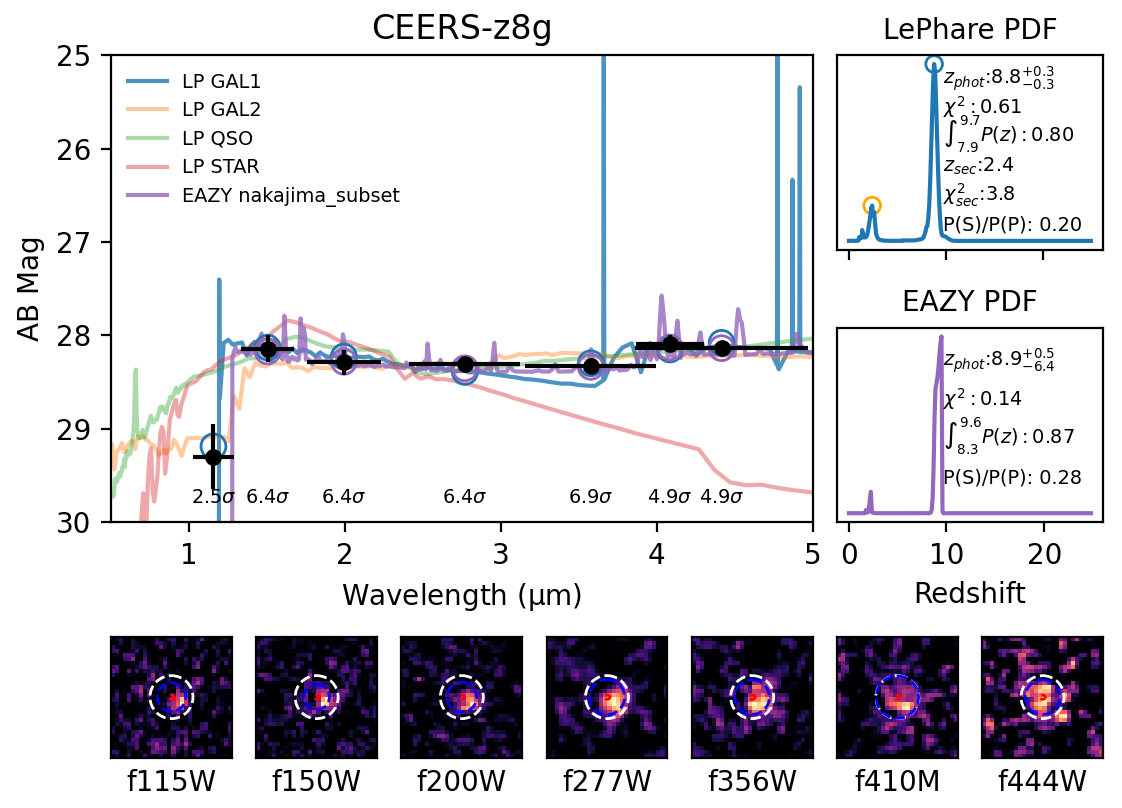}}\hfill
\subfloat{\includegraphics[width=\columnwidth]{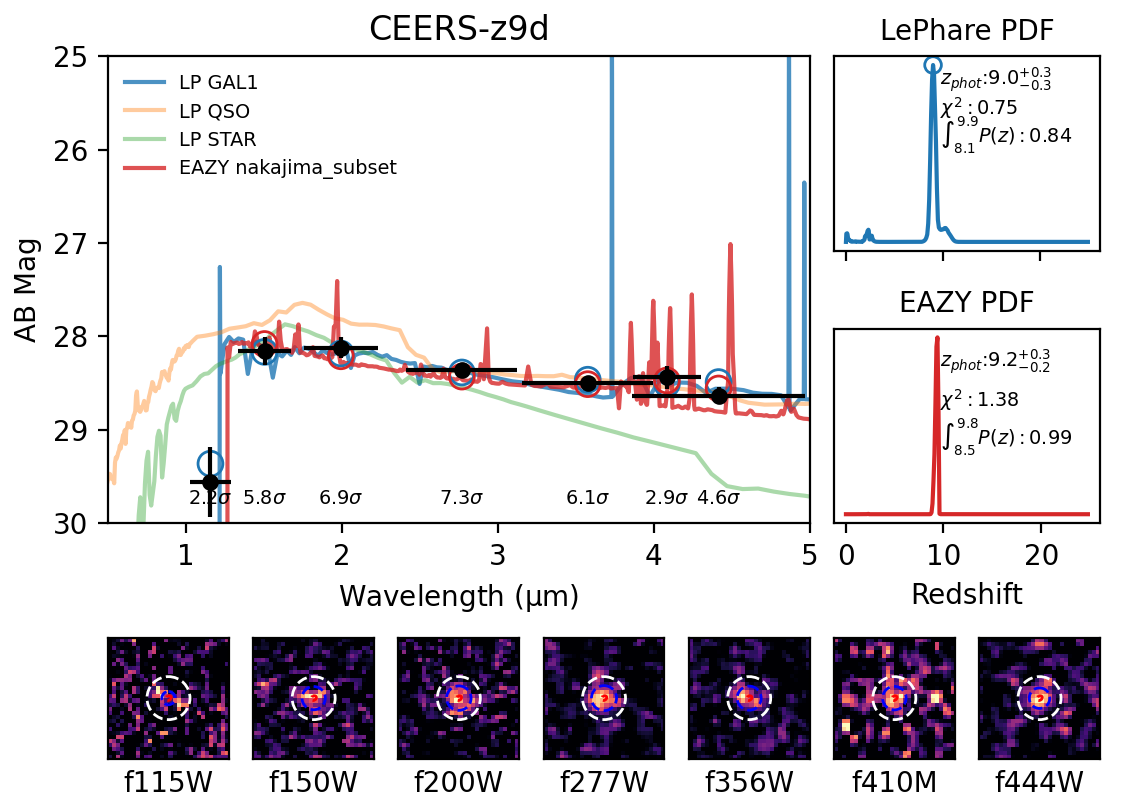}}\hfill
\caption{The other 4 candidate sources, excluding, S0723-z11c, which is discussed separately. }
\label{fig:SED_2}
\end{figure*}

\begin{table*}
    \centering
    \begin{tabular}{cccccccccc}
        \hline
         ID& R.A. & Dec & $z$ & $T_{bb}$ (K) & $\alpha$&$f_{AGN}$& $R_e$ (kpc) &$M_{UV}$ & $L_X$ (erg s$^{-1}$) \\
         \hline
         NEP-z7l& 260.86441 & 65.81502 & $8.2^{+0.5}_{-0.3}$& $2\times10^5$ &-2.0&$0.1 \pm 0.3$& $0.41 \pm 0.11$ & -19.26$^{+0.07}_{-0.07}$ &<10$^{43.0}$\\ 
         CEERS-z8i& 214.86438 & 52.77220 & $8.8^{+0.3}_{-0.3}$ & $2\times10^5$ & -1.6 & $0.8 \pm 0.3$ & 0.76 $\pm$ 0.12 &-19.16$^{+0.15}_{-0.18}$&<10$^{44.0}$\\ 
         NEP-z9b& 260.68414 & 65.72350 & $9.5^{+0.7}_{-0.7}$ &$2\times10^5$&-1.2&$0.8 \pm 0.3$& $<$ 0.624 & -19.50$^{+0.11}_{-0.13}$ & <10$^{43.2}$ \\
         NEP-z8ff& 260.80899 & 65.83811 & $8.7^{+0.2}_{-0.3}$ &$1\times10^5$&-1.2&$0.2 \pm 0.3$& 0.32 $\pm$ 0.06 & -19.70$^{+0.11}_{-0.13}$ & <10$^{43.1}$\\
         NEP-z6c& 260.81175 & 65.84165 & $6.5^{+0.2}_{-0.2}$&$5\times10^4$&-2.0&$0.1 \pm 0.3$& 1.32 $\pm$ 0.23 &-18.97$^{+0.12}_{-0.14}$&<10$^{42.8}$ \\
         CEERS-z8d& 215.01529 & 52.98497 & $8.8^{+0.2}_{-0.2}$& $1\times10^5$ &-2.0 & $0.2 \pm 0.3$ & 0.63 $\pm$ 0.07 & -19.58$^{+0.10}_{-0.11}$ & <10$^{44.0}$\\
         S0723-z11c& 110.69756 & -73.50050 & $11.9^{+0.3}_{-0.4}$ &$2\times10^5$ &-1.6 & $1.0 \pm 0.3$ & 0.63 $\pm$ 0.14 & -19.53$^{+0.10}_{-0.10}$ & -- \\
         CEERS-z8g& 214.96892 & 52.87178 & $8.8^{+0.3}_{-0.3}$ &$1\times10^5$&-1.2&$0.3 \pm 0.3$& 0.56 $\pm$ 0.10 &-19.14$^{+0.14}_{-0.16}$& <10$^{44.0}$\\
         CEERS-z9d& 214.87038 & 52.80202 & $9.0^{+0.3}_{-0.3}$ &$2\times10^5$&-1.2&$0.5 \pm 0.3$& $<$ 0.647 &-19.15$^{+0.14}_{-0.16}$&<10$^{44.0}$ \\
         CEERS 1019&  215.03539 & 52.89066 & 8.68* &--&--&$0.1 \pm 0.3$& $<$ 0.662  &-22.07$^{+0.05}_{-0.05}$&<$10^{44.2}$\\
         \hline
    \end{tabular}
    \caption{A table summarizing the basic properties of our candidate sources. The first three columns contain the catalog ID, separated from the field name by a dash, and right ascension and declination coordinates. Column 4 contains the photometric redshifts given by LePhare for our sources and the spectroscopic redshift for CEERS 1019, the uncertainty of which is not given due to being orders of magnitude lower than others. Columns 5 and 6 list big-bump temperatures and UV slopes given by the Nakajima template. Column 7, contains the best fit AGN fraction given by CIGALE. Radii measured in kpc are given in column 8. For sources best fit with a combined or Sersic model, this is taken from the half light radius of the extended Sersic profile, for those best fit by a PSF the radius is an upper limit given by the PSF FWHM. The last two columns contain absolute UV magnitudes, the one for CEERS 1019 was taken from literature, and X-ray luminosity limits in the 0.5 - 10 keV band for CEERS and NEP sources respectively.}
    \label{tab:sources}
\end{table*}

\subsection{X-ray and radio limits}

We check if any of our candidate sources present in the NEP and CEERS fields have X-ray detections by matching our final candidate catalog with Chandra deep field point source catalogs from AEGIS-X survey of the Extended Groth Strip, overlapping the CEERS field \cite{AEGIS-X}, and a deep X-ray survey of the JWST NEP field.  The matching was carried out using 0.3 arcsec matching radii. However, none of our sources in CEERS and NEP fields appear to have X-ray detections in Chandra data. Thus, we use it to estimate upper limits on the full band (0.5 - 10 keV) X-ray luminosity of our sources.

For the AEGIS-X data, we take the 50\% completeness limit of $1.30 \times 10^{-15}$ erg s$^{-1}$ cm$^{-2}$ from \cite{AEGIS-X} as our limiting flux. For the TDF survey we were able to determine a 3$\sigma$ detection limit by checking the catalog for the faintest sources that were detected at 3$\sigma$ significance. This came out to $6\times10^{-6}$~cps, using a conversion factor of 1~cps = $2.842\times10^{-11}$ erg s$^{-1}$ cm$^{-2}$, this resulted in an upper limit of $1.7\times10^{-16}$ erg s$^{-1}$ for sources in the NEP survey fields. It should be noted that the X-ray catalog for the NEP field did not have completeness estimates at the time of writing, thus this value may be an underestimate.   The calculated X-ray luminosity limits are of order $10^{43}$ - $10^{44}$~erg~ s$^{-1}$. This places our sources at or below the characteristic X-ray luminosity of $\sim$10$^{44}$~erg~ s$^{-1}$ for AGN at z = 4 - 5 \citep{XLF1, XLF2}.  However, our sources have low inferred stellar masses, so we probably would not expect the AGN to belong on the bright end of the luminosity function. 

We check for radio detections by matching our candidates in the NEP field with the Very Large Array (VLA) catalog for the same field, described by \cite{VLA_rad}. Using 2 arcsec matching radii, as expected, no matches were found between our candidate sources, thus giving limiting fluxes of 10~$\mu$Jy for all candidates in the NEP field, based on the flux cutoff in \cite{VLA_rad}.

\subsection{Near-infrared colors}

In order to compare the photometry of our selected candidates with theoretical predictions for DCBHs, we adopt two sets of NIR color cuts. The first one consists of the 90\% purity cuts for an AGN number fraction ($n_{AGN}$) of 25\% from Table 1 of \cite{Pop_3_agn}, which were tailored to identify low mass BHs at $7<z<10$ accreting at an Eddington ratio of >0.1. The second set was adopted from \cite{OBG} and was derived for a hypothetical class of obese black hole galaxies (OBG) at  $z \sim 9$, which form after a DCBH acquires a stellar emission component. 

\begin{figure}
    \centering
    \includegraphics[width=0.9\columnwidth]{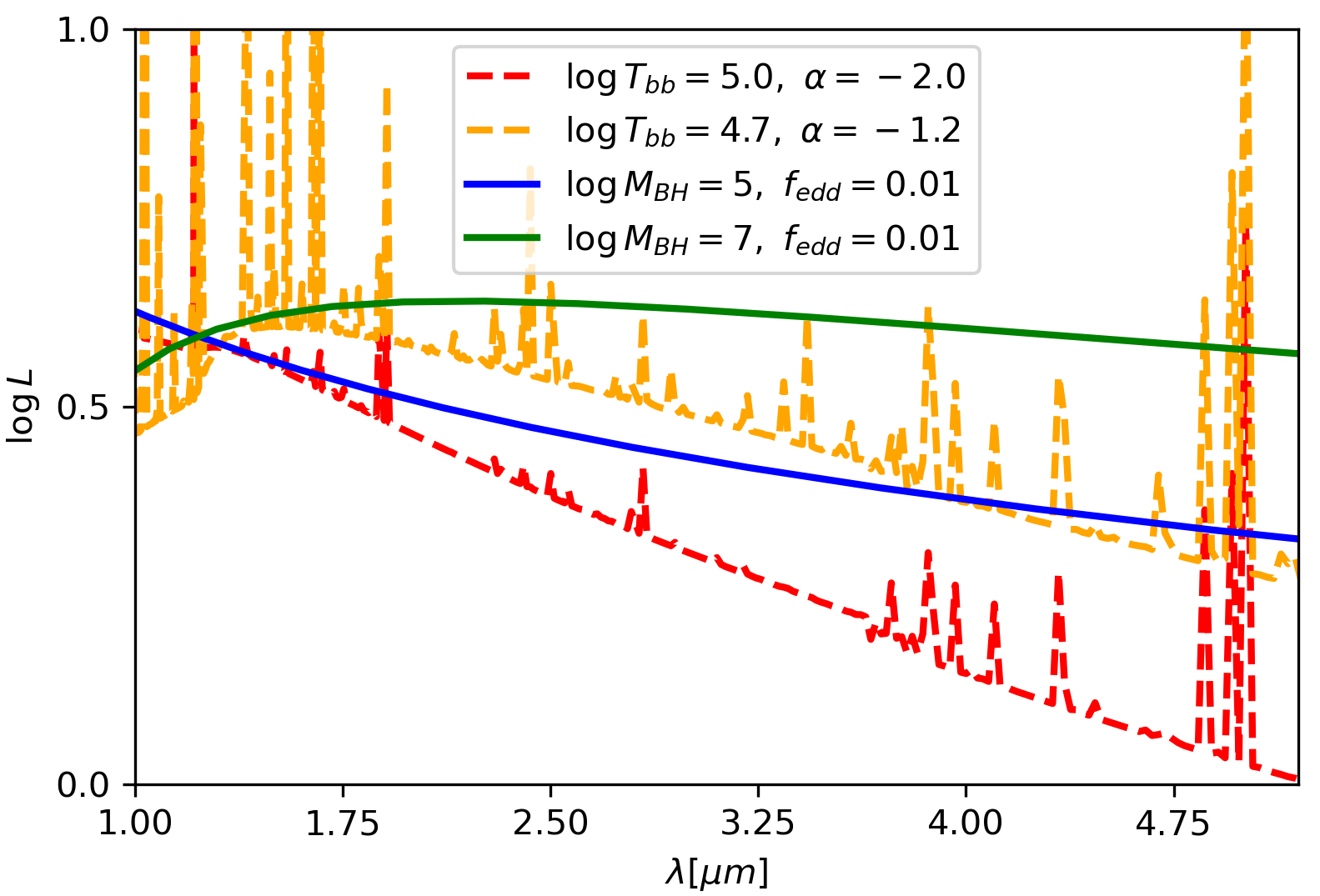}
    \caption{A comparison between the Nakajima SEDs (dashed lines) and AGN continuum SED models from \protect\cite{AGN_SED_param} (solid lines). The y axis is scaled to the same continuum luminosity, all SEDs were redshifted to $z=9$.}
    \label{fig:SED_comp}
\end{figure}

\begin{figure}
    \centering
    \includegraphics[width=8.5cm]{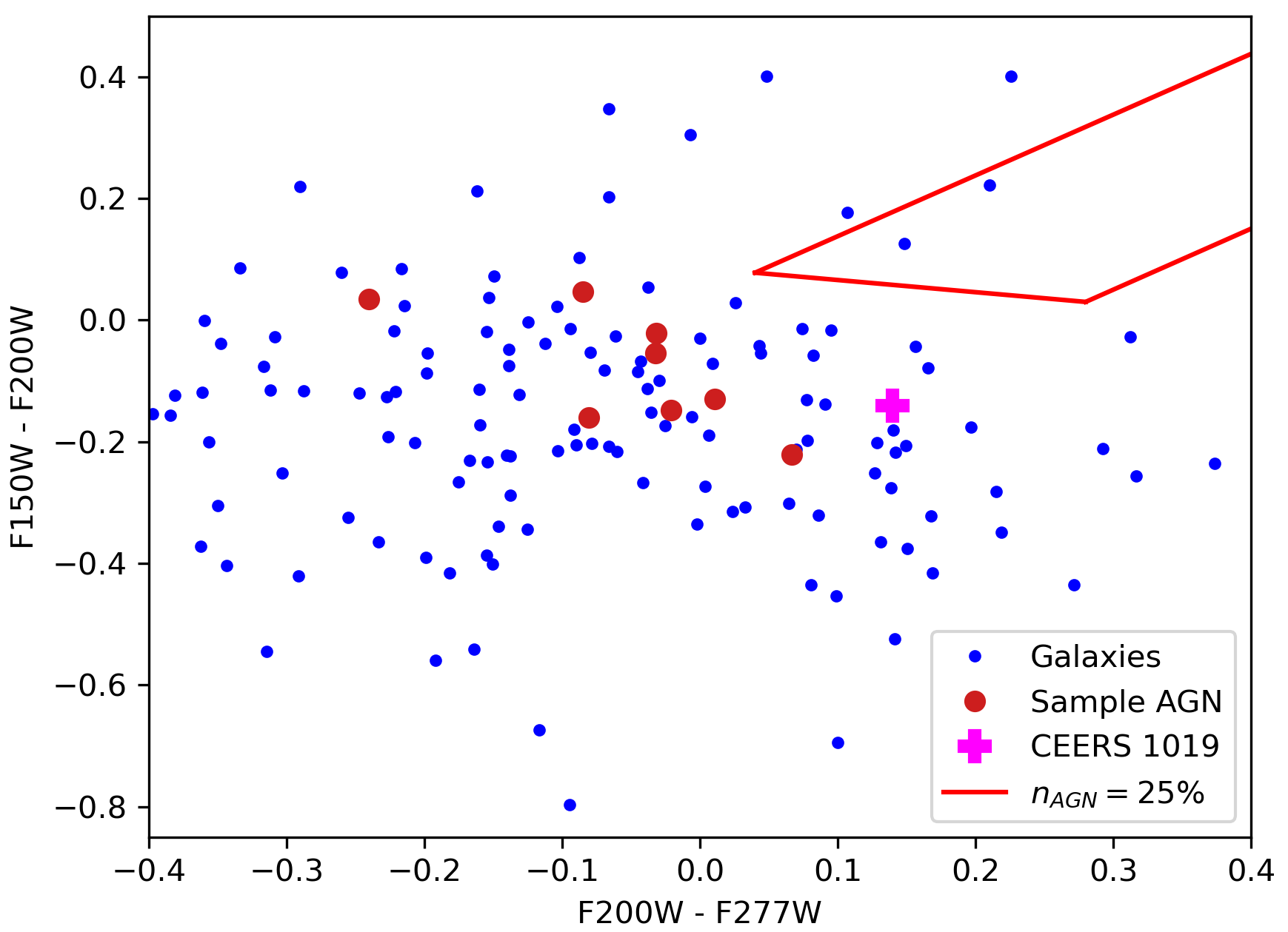}
    \caption{A plot of rest frame UV colors for our selected AGN sources, the remaining galaxies in our sample and CEERS 1019 which is a spectroscopiclly confirmed AGN at this redshift range \protect\cite{Larson2022}. The red lines denote the 90\% purity selection region for AGN sources, assuming 25\% of all galaxies host an AGN, given by  \protect\cite{Pop_3_agn}. S0723-z11c is excluded from this plot as it starts to drop out in F150W.  As can be seen we do not find that our sources are found within this region of the plot.  }
    \label{fig:goulding_cuts}
\end{figure}

Computing the colors using aperture corrected SExtractor magnitudes in each filter we found that our candidate sources have marginally flatter SEDs than the rest of the high-z galaxy sample, however, the overall colour difference is not substantial, as can be seen in \autoref{fig:goulding_cuts}. This same figure also shows that our sources are significantly bluer than the red predictions from \cite{Pop_3_agn, 2016MNRAS_directcollapse}, likely due to differing assumed SED sets. The key difference seems to stem from \cite{Pop_3_agn} assuming an $\alpha = -0.79$, derived by \cite{shields_slope} from low redshift narrow-line Seyfert 1 galaxies. We make a further comparison of the Nakajima SEDs with models used in \cite{AGN_SED_param}, which result in similar predicted colors to \cite{Pop_3_agn}. These models, hereafter referred to as the Volonteri set, are explicitly parametrized by the black hole mass ($M_{BH}$) and the Eddington ratio ($f_{edd}$) and do not include nebular emission lines, unlike the Nakajima set. A comparison between the bluest and reddest SEDs possible from both model sets in the considered wavelength range is provided in \autoref{fig:SED_comp}. It can be seen from the figure that running the Volonteri models with lower $M_{BH}$ results in bluer continuum shapes, however, the overall range of apparent slopes of Volonteri models is significantly redder than that of the Nakajima set. The likely reason for this is that the Volonteri models assume an $\alpha = -0.5$, following \cite{Thomas2016}. These slopes differ significantly from the steeper values assumed by the Nakajima model, following \cite{Elvis2002} results for low redshift quasars. Thus a possible reason for our objects not matching the \cite{Pop_3_agn} color cuts is the differing assumptions of the underlying SED models.

It should also be noted, however, that the CEERS 1019 source is likewise not significantly differentiated from either the high-z galaxies or our AGN sample in the \cite{Pop_3_agn} color space. The GN-z11 source, while not in our photometric sample, has also been reported to have a blue ($\beta = -2.26 \pm 0.1$) rest-frame UV slope \citep{GZ-z11}. These bluer than expected colors may also be partially attributed to some of our sources having a significant stellar component, as suggested by the AGN fractions given by CIGALE in \autoref{tab:sources}.

\begin{figure}
    \centering
    \includegraphics[width=8.5cm]{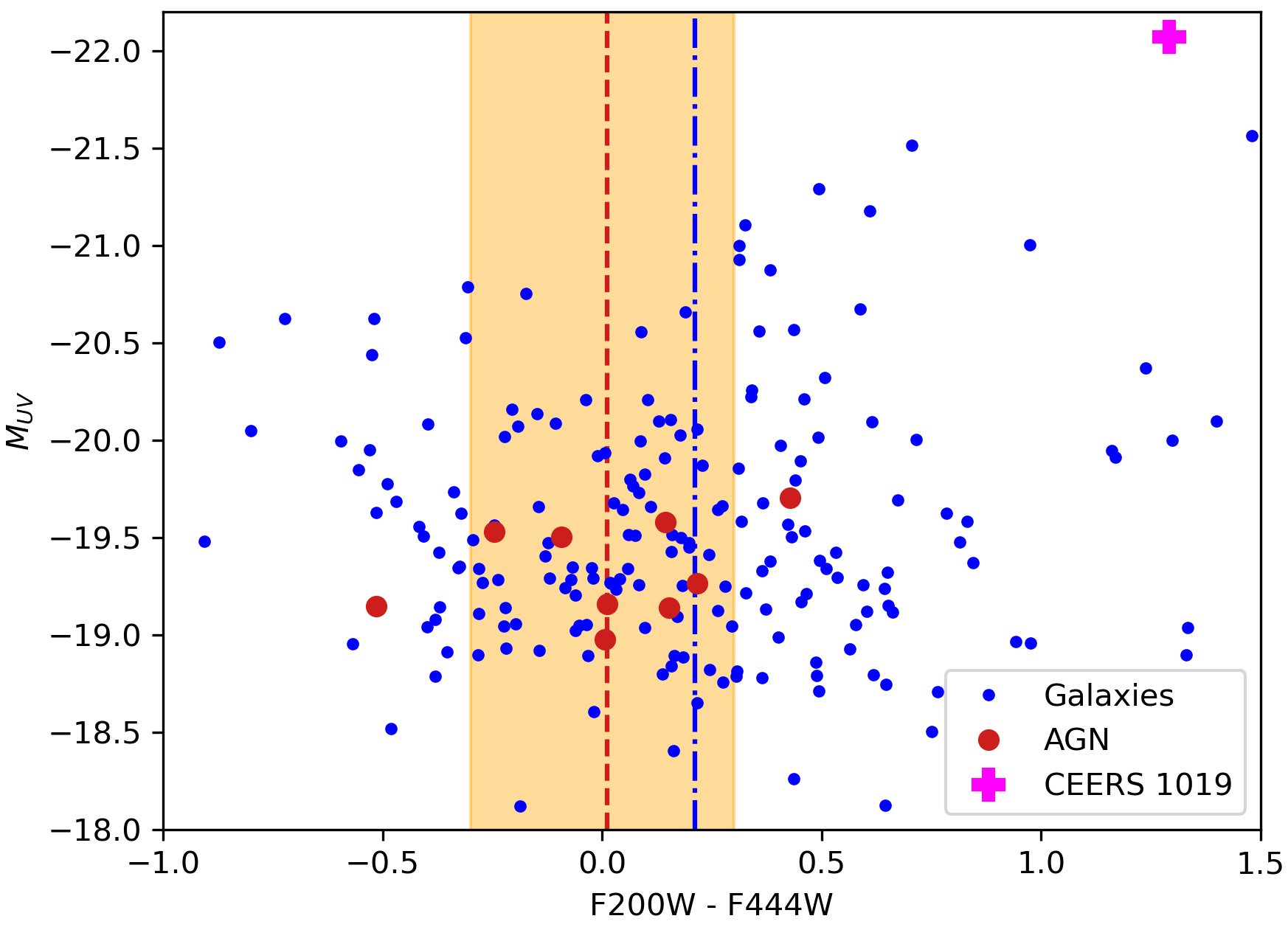}
    \caption{A comparison between our candidate AGN and the rest of the high-z galaxy sample plotted in terms of the (F200W - F444W) color and the measured UV absolute magnitude. The red dashed line denotes the average color for our candidate sample - the blue dot dashed line represents the average for galaxies that were not selected. The orange shaded region is the range of predicted OBG colors from \protect\cite{OBG}.}
    \label{fig:natarayan_cols}
\end{figure}

A comparison of the AGN candidates the rest of the galaxies in the (F200W - F444W) colour band (\autoref{fig:natarayan_cols})  shows the relative flatness of their spectra more clearly, with the average (F200W - F444W) color being close to 0, while the same average for the high-z galaxies is at $\sim 0.2$. The CEERS 1019 object appears redder in this figure, however, this is due to it possessing a strong OIII line above an otherwise flat continuum \citep{CEERS_z8_agn}. While the derived $M_{UV}$ values do not differentiate our sources from the rest of the sample, it should be noted that 7 out of 9 of our candidates lie in the region -0.3 < F200W - F444W < 0.3, in line with predictions from \cite{OBG}. Our magnitudes, however, are fainter by up to 1 mag than their predictions, assuming an optical bolometric correction $K_O = 5$ \citep{bolometric}. It should be noted that this correction was derived from AGN at $z< 4 $ and may not hold at the redshift range considered here. In general, much more work is needed to understand the SEDs and spectral characteristics of $z > 5$ AGN.

\subsection{Masses and start formation rates}
We adopt the star formation rates (SFR) from the parent sample of 214 sources. These SFRs were estimated by taking the average flux in the restframe 1450 and 1550 \r{A} wavelength range, using it to calculate the UV luminosity, which, after dust corrections from \cite{Meurer1999}, is converted into SFR using the procedure described in \cite{MadauDickinson2014}. Stellar masses were obtained by fitting the sample sources with {\tt{BAGPIPES}} \cite{bagpipes}.

\begin{figure}
    \centering
    \includegraphics[width=0.5\textwidth]{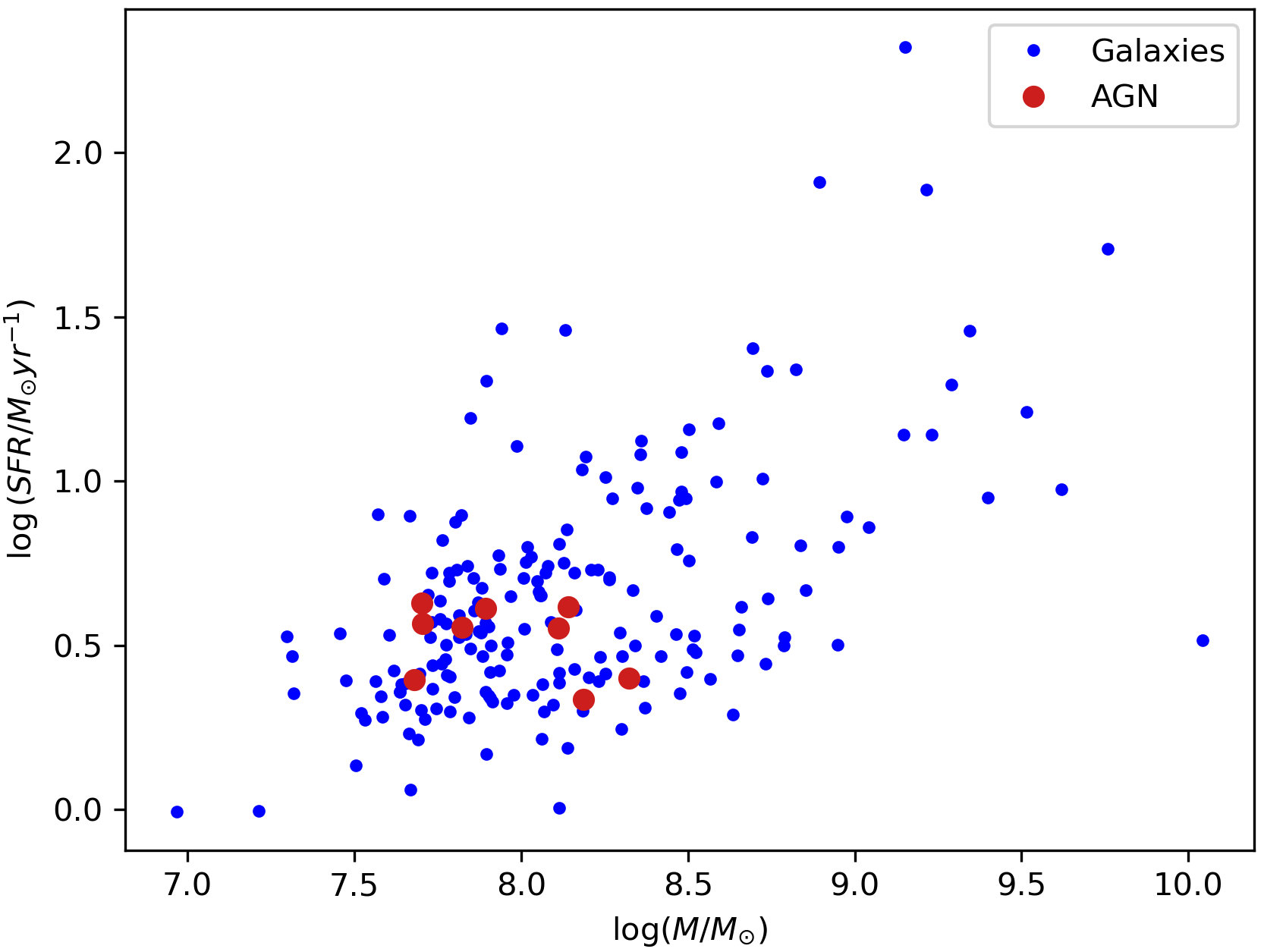}
    \caption{A plot comparing the star formation rates and stellar masses, inferred by assuming stellar emissions, of our candidate AGN and the remaining galaxy sample. As can be seen, our sources do not show up as significant outliers in this plot.}
    \label{fig:MassSFR}
\end{figure}

The above methods do not take into account potential AGN emissions, however, a comparison between our candidates and the parent sample may be useful in seeing if AGN may be efficiently identified by looking at outliers in an SFR versus stellar mass plot. Such a plot is presented in \autoref{fig:MassSFR}. As can be seen from the figure, calculating stellar masses and SFRs assuming purely stellar emission does not produce anomalous results for our candidates, likely due to their low UV luminosity. This low luminosity could be the result of low black hole masses and accretion rates. However, an intriguing possibility is that some of our AGN may be the high redshift counterparts of sources found by \cite{LabbeAGN}, which are characterized by a faint and relatively blue UV continuum and a bright, red rest-frame optical SED. At the redshifts considered in this paper the red continuum would mostly lie outside of the NIRCam range. Thus deep mid-infrared observations are required to check this hypothesis.

\subsection{An unusual object at z = 12}

\begin{figure}
    \centering
    \includegraphics[width=0.5\textwidth]{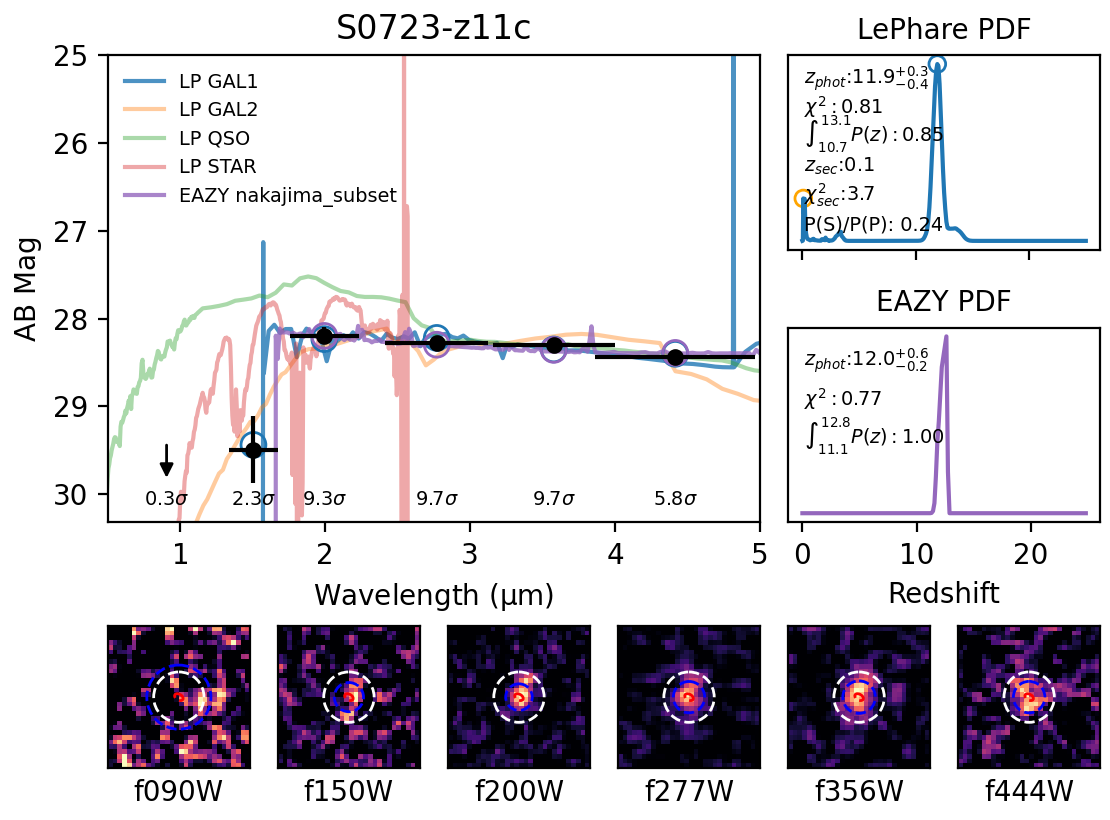}
    \caption{The best fit SEDs for the source S0723-z11c, our highest redshift AGN candidate. The plots are organized as in figure \ref{fig:SED_1}. }
    \label{fig:6286}
    \hfill
\end{figure}

\begin{figure}
    \centering
    \includegraphics[width=0.5\textwidth]{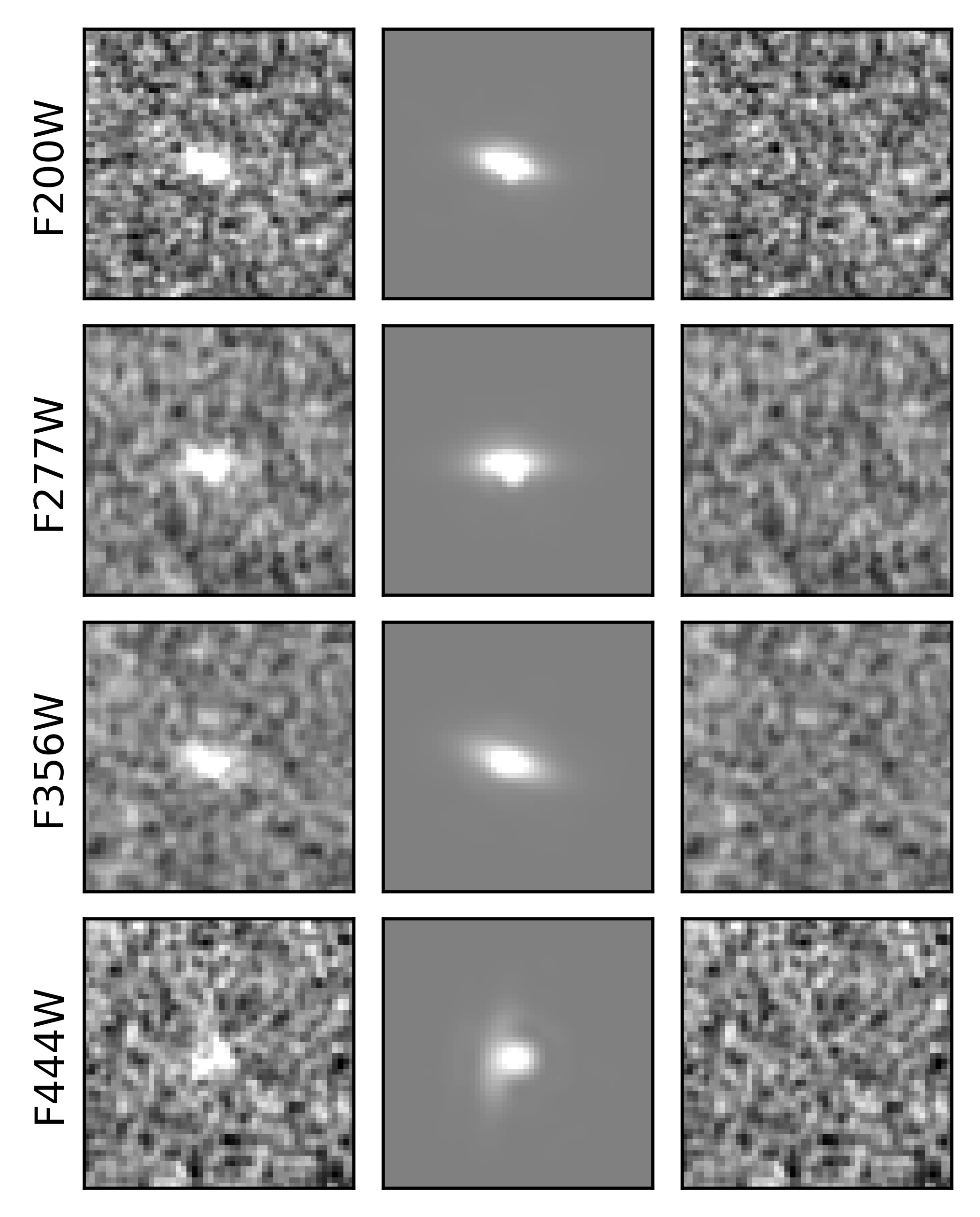}
    \caption{This figure shows the light in different wavebands for the unusual source S0723-z11c.  We show a comparison between the actual data image (left), models obtained by \galfit\ (centre), and the residual images (right), for S0723-z11c in the different bands in which the source is observed. Each cutout is 1.5'' x 1.5''. The source is fit using a combined Sersic and PSF model in all bands.}
    \label{fig:s0723_comp}
\end{figure}

In terms of individual sources,  S0723-z11c stands out as our candidate at the highest redshift of $\sim12$. As can be seen from \autoref{fig:6286}, LePhare galaxy models give similar performance to the Nakajima set in terms of $\chi^2$ values, however, the image cutouts presented in the same figure showcase a composite and complex nature of the source. The morphological fits in \autoref{fig:morph} identify the second component as a point source, contributing almost 40\% to the total emission in the F444W band. However, it is important to note that the apparent morphology changes somewhat drastically in this band with respect to others, as highlighted in \autoref{fig:s0723_comp}. In order to better understand the complex morphology of this source, we run {\tt{GALFIT}} across all bands in which the source had >5$\sigma$ detections (F200W, F277W, F356W and F444W), with the results summarized in \autoref{fig:s0723_comp}. As can be seen from the figure, the source in each band is best-fit by a combination of a Sersic profile and a PSF, with component locations being roughly consistent from F200W to F356W, with a rather abrupt location shift occurring in the F444W band. The consistent bimodal nature of this source along with the shift in the F444W band may point to a morphology disturbed by a merger event or a strong outflow.

A possible reason behind the abrupt nature of the shift between F356W and F444W band images is either line emission or a discontinuity in the continuum itself. Assuming a source redshift of $z = 12$, this emission feature should occur at rest-frame wavelengths of 300 - 383~nm. While this may be caused by either a NeV (3346~\r{A}, 3426~\r{A}) or OII (3727~\r{A}) doublets, the spatially extended nature of this emission suggests that it may be due to a Balmer break, which in turn would indicate the presence of evolved stellar populations in the object. However, observations probing redwards of the F444W band or a spectroscopic followup is required to truly confirm the nature of this discontinuity in emission patterns.

\section{Conclusions and Summary}\label{sec:conc}

In this paper we have identified a population of high redshift AGN candidates by utilizing a set photometric and morphological selection techniques.   The basic idea behind our method and paper is to find systems that are much better fit by AGN templates with an active galaxy, or black holes, and are consistent with having a small point source that dominates the light of the galaxy.  our methods is not meant to find complete samples of AGN or early black holes, but as a way to find the best candidates for further spectroscopy and detailed follow up.

Our parent samples originates from the EPOCHS sample of $z > 6.5$ galaxies whose luminosity function and selection is discussed in \citep[][]{UV_LF_methods}.  From this sample we refit the galaxies using a variety of model SEDs using the photometric redshift code EAZY.   We use this method to isolate most of the sample sources, however, other steps in our pipeline - a combination of analysis of relative performance between AGN and non-AGN models and morphological fitting was also utilised in removing weak candidates.  We are thus left with nine strong AGN sources that are likely emitting their light due to a central massive black hole.
 
We develop a new method of finding likely AGN through template fitting, such that we define a statistic $P_{AGN}$ that reveals how likely an object is better fit by an AGN spectrum rather than a star forming one.   It should also be noted that the reason for the $P_{AGN}$ limited performance in isolation may be its implicit reliance on the general properties of the $z>6$ AGN population being well known.  New AGN templates are needed at these highest redshifts which will eventually be developed with the availability of more JWST spectroscopy of these objects.   Our overall selection method may, however, provide a good way for searching for the highest redshift candidate AGN for spectroscopic followup using NIRSpec.

We find that the estimated AGN fraction in the interval of $6.5 < z < 12$ is $5\pm 1\%$. However, our investigation was strongly biased towards Type 1 AGN, due to the initial set of SED templates not accounting for dust extinction, and calibrated for purity rather than completeness, thus this result only establishes a lower bound, which is nevertheless consistent with theoretical predictions.

We also find that rest-frame UV photometry of our candidates suggest that color-color cuts alone may not be sufficient to differentiate AGN from other galaxies at high redshifts - with SED and morphological fitting in conjunction with deep X-ray and spectroscopic observations being necessary for robust identification. However, color-color cuts may still be useful as a way to pre-select potential candidates, as evidenced by a large fraction of our sources lining up with bluer colors predicted for OBGs.

The presence of JWST-detectable AGN sources out to z = 12 alone suggests evidence in favour of the Direct Collapse Black Hole model \citep{JWST_bhsearch, AGN_SED_param}, while the photometric properties of our sample suggest evidence in favour of the OBG stage of galaxy formation and thus a type of 'naked' black holes existing in the early Universe, however, follow up spectroscopy will be needed to confirm the nature of our objects and estimating their black hole masses in order to place more defined constraints on black hole seeding models.

\section*{Acknowledgements}

We gratefully acknowledge support from the ERC Advanced Investigator Grant EPOCHS (788113), as well as studentships from STFC. LF acknowledges financial support from Coordenação de Aperfeiçoamento de Pessoal de Nível Superior - Brazil (CAPES) in the form of a PhD studentship. DI acknowledges support by the European Research Council via ERC Consolidator Grant KETJU (no. 818930). CCL acknowledges support from the Royal Society under grant RGF/EA/181016. CT acknowledges funding from the Science and Technology Facilities Council (STFC). This work is based on observations made with the NASA/ESA \textit{Hubble Space Telescope} (HST) and NASA/ESA/CSA \textit{James Webb Space Telescope} (JWST) obtained from the \texttt{Mikulski Archive for Space Telescopes} (\texttt{MAST}) at the \textit{Space Telescope Science Institute} (STScI), which is operated by the Association of Universities for Research in Astronomy, Inc., under NASA contract NAS 5-03127 for JWST, and NAS 5–26555 for HST. The Early Release Observations and associated materials were developed, executed, and compiled by the ERO production team: Hannah Braun, Claire Blome, Matthew Brown, Margaret Carruthers, Dan Coe, Joseph DePasquale, Nestor Espinoza, Macarena Garcia Marin, Karl Gordon, Alaina Henry, Leah Hustak, Andi James, Ann Jenkins, Anton Koekemoer, Stephanie LaMassa, David Law, Alexandra Lockwood, Amaya Moro-Martin, Susan Mullally, Alyssa Pagan, Dani Player, Klaus Pontoppidan, Charles Proffitt, Christine Pulliam, Leah Ramsay, Swara Ravindranath, Neill Reid, Massimo Robberto, Elena Sabbi, Leonardo Ubeda. The EROs were also made possible by the foundational efforts and support from the JWST instruments, STScI planning and scheduling, and Data Management teams. The effort of CEERS, NGDEEP and GLASS teams in making their data public is hereby acknowledged. We would also like to thank Adi Zitrin, Rachana Bhatawdekar and Nimish Hathi for their timely and useful comments.

\section*{Data Availability}
Some of the data data underlying this article is made available by \cite{UV_LF_methods}. The remainder of the data set will be released together with Conselice et al. 2023. The catalogues of the sample discussed herein may be acquired by contacting the corresponding author.



\bibliographystyle{mnras}
\bibliography{references} 

\begin{thebibliography}{}
\makeatletter
\relax
\def\mn@urlcharsother{\let\do\@makeother \do\$\do\&\do\#\do\^\do\_\do\%\do\~}
\def\mn@doi{\begingroup\mn@urlcharsother \@ifnextchar [ {\mn@doi@}
  {\mn@doi@[]}}
\def\mn@doi@[#1]#2{\def\@tempa{#1}\ifx\@tempa\@empty \href
  {http://dx.doi.org/#2} {doi:#2}\else \href {http://dx.doi.org/#2} {#1}\fi
  \endgroup}
\def\mn@eprint#1#2{\mn@eprint@#1:#2::\@nil}
\def\mn@eprint@arXiv#1{\href {http://arxiv.org/abs/#1} {{\tt arXiv:#1}}}
\def\mn@eprint@dblp#1{\href {http://dblp.uni-trier.de/rec/bibtex/#1.xml}
  {dblp:#1}}
\def\mn@eprint@#1:#2:#3:#4\@nil{\def\@tempa {#1}\def\@tempb {#2}\def\@tempc
  {#3}\ifx \@tempc \@empty \let \@tempc \@tempb \let \@tempb \@tempa \fi \ifx
  \@tempb \@empty \def\@tempb {arXiv}\fi \@ifundefined
  {mn@eprint@\@tempb}{\@tempb:\@tempc}{\expandafter \expandafter \csname
  mn@eprint@\@tempb\endcsname \expandafter{\@tempc}}}

\bibitem[\protect\citeauthoryear{{Adams} et~al.,}{{Adams}
  et~al.}{2023a}]{UV_LF_methods}
{Adams} N.~J.,  et~al., 2023a, arXiv e-prints, \href
  {https://ui.adsabs.harvard.edu/abs/2023arXiv230413721A} {p. arXiv:2304.13721}

\bibitem[\protect\citeauthoryear{{Adams} et~al.,}{{Adams}
  et~al.}{2023b}]{2023MNRAS.518.4755A}
{Adams} N.~J.,  et~al., 2023b, \mn@doi [\mnras] {10.1093/mnras/stac3347}, \href
  {https://ui.adsabs.harvard.edu/abs/2023MNRAS.518.4755A} {518, 4755}

\bibitem[\protect\citeauthoryear{{Aird}, {Coil}, {Georgakakis}, {Nandra},
  {Barro}  \& {P{\'e}rez-Gonz{\'a}lez}}{{Aird} et~al.}{2015}]{XLF2}
{Aird} J.,  {Coil} A.~L.,  {Georgakakis} A.,  {Nandra} K.,  {Barro} G.,
  {P{\'e}rez-Gonz{\'a}lez} P.~G.,  2015, \mn@doi [\mnras]
  {10.1093/mnras/stv1062}, \href
  {https://ui.adsabs.harvard.edu/abs/2015MNRAS.451.1892A} {451, 1892}

\bibitem[\protect\citeauthoryear{{Bagley} et~al.,}{{Bagley}
  et~al.}{2022}]{Bagley2022}
{Bagley} M.~B.,  et~al., 2022, \mn@doi [arXiv e-prints]
  {10.48550/arXiv.2211.02495}, \href
  {https://ui.adsabs.harvard.edu/abs/2022arXiv221102495B} {p. arXiv:2211.02495}

\bibitem[\protect\citeauthoryear{{Bagley} et~al.,}{{Bagley}
  et~al.}{2023}]{Bagley2023}
{Bagley} M.~B.,  et~al., 2023, \mn@doi [arXiv e-prints]
  {10.48550/arXiv.2302.05466}, \href
  {https://ui.adsabs.harvard.edu/abs/2023arXiv230205466B} {p. arXiv:2302.05466}

\bibitem[\protect\citeauthoryear{Bellovary, Volonteri, Governato, Shen, Quinn
  \& Wadsley}{Bellovary et~al.}{2011}]{Bellovary_2011}
Bellovary J.,  Volonteri M.,  Governato F.,  Shen S.,  Quinn T.,   Wadsley J.,
  2011, \mn@doi [\apj] {10.1088/0004-637X/742/1/13}, 742, 13

\bibitem[\protect\citeauthoryear{{Bertin} \& {Arnouts}}{{Bertin} \&
  {Arnouts}}{1996}]{sextractor}
{Bertin} E.,  {Arnouts} S.,  1996, \mn@doi [\aaps] {10.1051/aas:1996164}, \href
  {https://ui.adsabs.harvard.edu/abs/1996A&AS..117..393B} {117, 393}

\bibitem[\protect\citeauthoryear{{Brammer}, {van Dokkum}  \& {Coppi}}{{Brammer}
  et~al.}{2008}]{EAZY}
{Brammer} G.~B.,  {van Dokkum} P.~G.,   {Coppi} P.,  2008, \mn@doi [\apj]
  {10.1086/591786}, \href
  {https://ui.adsabs.harvard.edu/abs/2008ApJ...686.1503B} {686, 1503}

\bibitem[\protect\citeauthoryear{Bruzual \& Charlot}{Bruzual \&
  Charlot}{2003}]{Bruzual2003}
Bruzual G.,  Charlot S.,  2003, \mn@doi [\mnras]
  {10.1046/j.1365-8711.2003.06897.x}, 344, 1000

\bibitem[\protect\citeauthoryear{{Bushouse} et~al.,}{{Bushouse}
  et~al.}{2022}]{JWST_pipeline}
{Bushouse} H.,  et~al., 2022, {JWST Calibration Pipeline}, Zenodo,
  \mn@doi{10.5281/zenodo.7325378}

\bibitem[\protect\citeauthoryear{{Carnall}, {McLure}, {Dunlop}  \&
  {Dav{\'e}}}{{Carnall} et~al.}{2018}]{bagpipes}
{Carnall} A.~C.,  {McLure} R.~J.,  {Dunlop} J.~S.,   {Dav{\'e}} R.,  2018,
  \mn@doi [\mnras] {10.1093/mnras/sty2169}, \href
  {https://ui.adsabs.harvard.edu/abs/2018MNRAS.480.4379C} {480, 4379}

\bibitem[\protect\citeauthoryear{Conroy \& Gunn}{Conroy \&
  Gunn}{2010}]{conroy2010fsps}
Conroy C.,  Gunn J.~E.,  2010, Astrophysics Source Code Library, pp ascl--1010

\bibitem[\protect\citeauthoryear{{Constantin} \& {Shields}}{{Constantin} \&
  {Shields}}{2003}]{shields_slope}
{Constantin} A.,  {Shields} J.~C.,  2003, \mn@doi [\pasp] {10.1086/374724},
  \href {https://ui.adsabs.harvard.edu/abs/2003PASP..115..592C} {115, 592}

\bibitem[\protect\citeauthoryear{Dewsnap, Barmby, Gallagher, Urry, Ghosh  \&
  Powell}{Dewsnap et~al.}{2023}]{dewsnap2023}
Dewsnap C.,  Barmby P.,  Gallagher S.~C.,  Urry C.~M.,  Ghosh A.,   Powell
  M.~C.,  2023, \mn@doi [The Astrophysical Journal]
  {https://doi.org/10.3847/1538-4357/ac9400}, 944, 137

\bibitem[\protect\citeauthoryear{{Duras} et~al.,}{{Duras}
  et~al.}{2020}]{bolometric}
{Duras} F.,  et~al., 2020, \mn@doi [\aap] {10.1051/0004-6361/201936817}, \href
  {https://ui.adsabs.harvard.edu/abs/2020A&A...636A..73D} {636, A73}

\bibitem[\protect\citeauthoryear{{Elvis}, {Risaliti}  \& {Zamorani}}{{Elvis}
  et~al.}{2002}]{Elvis2002}
{Elvis} M.,  {Risaliti} G.,   {Zamorani} G.,  2002, \mn@doi [\apjl]
  {10.1086/339197}, \href
  {https://ui.adsabs.harvard.edu/abs/2002ApJ...565L..75E} {565, L75}

\bibitem[\protect\citeauthoryear{{Ferreira} et~al.,}{{Ferreira}
  et~al.}{2022}]{2022ApJ...938L...2F}
{Ferreira} L.,  et~al., 2022, \mn@doi [\apjl] {10.3847/2041-8213/ac947c}, \href
  {https://ui.adsabs.harvard.edu/abs/2022ApJ...938L...2F} {938, L2}

\bibitem[\protect\citeauthoryear{{Finkelstein} et~al.,}{{Finkelstein}
  et~al.}{2022}]{2022arXiv221105792F}
{Finkelstein} S.~L.,  et~al., 2022, arXiv e-prints, \href
  {https://ui.adsabs.harvard.edu/abs/2022arXiv221105792F} {p. arXiv:2211.05792}

\bibitem[\protect\citeauthoryear{{Fotopoulou} et~al.,}{{Fotopoulou}
  et~al.}{2016}]{XLF1}
{Fotopoulou} S.,  et~al., 2016, \mn@doi [\aap] {10.1051/0004-6361/201424763},
  \href {https://ui.adsabs.harvard.edu/abs/2016A&A...587A.142F} {587, A142}

\bibitem[\protect\citeauthoryear{{Fudamoto}, {Inoue}  \& {Sugahara}}{{Fudamoto}
  et~al.}{2022}]{2022ApJ...938L..24F}
{Fudamoto} Y.,  {Inoue} A.~K.,   {Sugahara} Y.,  2022, \mn@doi [\apjl]
  {10.3847/2041-8213/ac982b}, \href
  {https://ui.adsabs.harvard.edu/abs/2022ApJ...938L..24F} {938, L24}

\bibitem[\protect\citeauthoryear{Furtak et~al.,}{Furtak
  et~al.}{2023}]{furtak2023jwst}
Furtak L.~J.,  et~al., 2023, JWST UNCOVER: Extremely red and compact object
  at$z_{\mathrm{phot}}\simeq7.6$ triply imaged by Abell 2744 (\mn@eprint
  {arXiv} {2212.10531})

\bibitem[\protect\citeauthoryear{{Gao}, {Yoshida}, {Abel}, {Frenk}, {Jenkins}
  \& {Springel}}{{Gao} et~al.}{2007}]{first_pop3}
{Gao} L.,  {Yoshida} N.,  {Abel} T.,  {Frenk} C.~S.,  {Jenkins} A.,
  {Springel} V.,  2007, \mn@doi [\mnras] {10.1111/j.1365-2966.2007.11814.x},
  \href {https://ui.adsabs.harvard.edu/abs/2007MNRAS.378..449G} {378, 449}

\bibitem[\protect\citeauthoryear{{Goulding} \& {Greene}}{{Goulding} \&
  {Greene}}{2022}]{Pop_3_agn}
{Goulding} A.~D.,  {Greene} J.~E.,  2022, \mn@doi [\apjl]
  {10.3847/2041-8213/ac9614}, \href
  {https://ui.adsabs.harvard.edu/abs/2022ApJ...938L...9G} {938, L9}

\bibitem[\protect\citeauthoryear{{Grogin} et~al.,}{{Grogin}
  et~al.}{2011}]{2011ApJS..197...35G}
{Grogin} N.~A.,  et~al., 2011, \mn@doi [\apjs] {10.1088/0067-0049/197/2/35},
  \href {https://ui.adsabs.harvard.edu/abs/2011ApJS..197...35G} {197, 35}

\bibitem[\protect\citeauthoryear{{Harikane} et~al.,}{{Harikane}
  et~al.}{2023}]{Harikane2023_agn}
{Harikane} Y.,  et~al., 2023, \mn@doi [arXiv e-prints]
  {10.48550/arXiv.2303.11946}, \href
  {https://ui.adsabs.harvard.edu/abs/2023arXiv230311946H} {p. arXiv:2303.11946}

\bibitem[\protect\citeauthoryear{{Hyun} et~al.,}{{Hyun} et~al.}{2023}]{VLA_rad}
{Hyun} M.,  et~al., 2023, \mn@doi [\apjs] {10.3847/1538-4365/ac9bf4}, \href
  {https://ui.adsabs.harvard.edu/abs/2023ApJS..264...19H} {264, 19}

\bibitem[\protect\citeauthoryear{{Kocevski} et~al.,}{{Kocevski}
  et~al.}{2023}]{hidden_monsters}
{Kocevski} D.~D.,  et~al., 2023, \mn@doi [arXiv e-prints]
  {10.48550/arXiv.2302.00012}, \href
  {https://ui.adsabs.harvard.edu/abs/2023arXiv230200012K} {p. arXiv:2302.00012}

\bibitem[\protect\citeauthoryear{{Koekemoer} et~al.,}{{Koekemoer}
  et~al.}{2011}]{2011ApJS..197...36K}
{Koekemoer} A.~M.,  et~al., 2011, \mn@doi [\apjs] {10.1088/0067-0049/197/2/36},
  \href {https://ui.adsabs.harvard.edu/abs/2011ApJS..197...36K} {197, 36}

\bibitem[\protect\citeauthoryear{{Labbe} et~al.,}{{Labbe}
  et~al.}{2023}]{LabbeAGN}
{Labbe} I.,  et~al., 2023, \mn@doi [arXiv e-prints]
  {10.48550/arXiv.2306.07320}, \href
  {https://ui.adsabs.harvard.edu/abs/2023arXiv230607320L} {p. arXiv:2306.07320}

\bibitem[\protect\citeauthoryear{{Laird} et~al.,}{{Laird}
  et~al.}{2009}]{AEGIS-X}
{Laird} E.~S.,  et~al., 2009, \mn@doi [\apjs] {10.1088/0067-0049/180/1/102},
  \href {https://ui.adsabs.harvard.edu/abs/2009ApJS..180..102L} {180, 102}

\bibitem[\protect\citeauthoryear{Larson et~al.,}{Larson
  et~al.}{2022}]{Larson2022}
Larson R.~L.,  et~al., 2022, Spectral Templates Optimal for Selecting Galaxies
  at z$>$8 with JWST, \mn@doi{10.48550/ARXIV.2211.10035}, \url
  {https://arxiv.org/abs/2211.10035}

\bibitem[\protect\citeauthoryear{{Larson} et~al.,}{{Larson}
  et~al.}{2023}]{CEERS_z8_agn}
{Larson} R.~L.,  et~al., 2023, \mn@doi [arXiv e-prints]
  {10.48550/arXiv.2303.08918}, \href
  {https://ui.adsabs.harvard.edu/abs/2023arXiv230308918L} {p. arXiv:2303.08918}

\bibitem[\protect\citeauthoryear{{Madau}}{{Madau}}{1995}]{Madau1995}
{Madau} P.,  1995, \mn@doi [\apj] {10.1086/175332}, \href
  {https://ui.adsabs.harvard.edu/abs/1995ApJ...441...18M} {441, 18}

\bibitem[\protect\citeauthoryear{{Madau} \& {Dickinson}}{{Madau} \&
  {Dickinson}}{2014}]{MadauDickinson2014}
{Madau} P.,  {Dickinson} M.,  2014, \mn@doi [\araa]
  {10.1146/annurev-astro-081811-125615}, \href
  {https://ui.adsabs.harvard.edu/abs/2014ARA&A..52..415M} {52, 415}

\bibitem[\protect\citeauthoryear{{Maiolino} et~al.,}{{Maiolino}
  et~al.}{2023}]{GZ-z11}
{Maiolino} R.,  et~al., 2023, \mn@doi [arXiv e-prints]
  {10.48550/arXiv.2305.12492}, \href
  {https://ui.adsabs.harvard.edu/abs/2023arXiv230512492M} {p. arXiv:2305.12492}

\bibitem[\protect\citeauthoryear{Matthee et~al.,}{Matthee
  et~al.}{2023}]{matthee2023little}
Matthee J.,  et~al., 2023, Little Red Dots: an abundant population of faint AGN
  at $z\sim5$ revealed by the EIGER and FRESCO JWST surveys (\mn@eprint {arXiv}
  {2306.05448})

\bibitem[\protect\citeauthoryear{{Meurer}, {Heckman}  \& {Calzetti}}{{Meurer}
  et~al.}{1999}]{Meurer1999}
{Meurer} G.~R.,  {Heckman} T.~M.,   {Calzetti} D.,  1999, \mn@doi [\apj]
  {10.1086/307523}, \href
  {https://ui.adsabs.harvard.edu/abs/1999ApJ...521...64M} {521, 64}

\bibitem[\protect\citeauthoryear{{Naidu} et~al.,}{{Naidu}
  et~al.}{2022a}]{Naidu2022b}
{Naidu} R.~P.,  et~al., 2022a, \mn@doi [arXiv e-prints]
  {10.48550/arXiv.2208.02794}, \href
  {https://ui.adsabs.harvard.edu/abs/2022arXiv220802794N} {p. arXiv:2208.02794}

\bibitem[\protect\citeauthoryear{{Naidu} et~al.,}{{Naidu}
  et~al.}{2022b}]{2022ApJ...940L..14N}
{Naidu} R.~P.,  et~al., 2022b, \mn@doi [\apjl] {10.3847/2041-8213/ac9b22},
  \href {https://ui.adsabs.harvard.edu/abs/2022ApJ...940L..14N} {940, L14}

\bibitem[\protect\citeauthoryear{{Nakajima} \& {Maiolino}}{{Nakajima} \&
  {Maiolino}}{2022}]{DCBH_templates}
{Nakajima} K.,  {Maiolino} R.,  2022, \mn@doi [\mnras]
  {10.1093/mnras/stac1242}, \href
  {https://ui.adsabs.harvard.edu/abs/2022MNRAS.513.5134N} {513, 5134}

\bibitem[\protect\citeauthoryear{{Natarajan}, {Pacucci}, {Ferrara}, {Agarwal},
  {Ricarte}, {Zackrisson}  \& {Cappelluti}}{{Natarajan} et~al.}{2017}]{OBG}
{Natarajan} P.,  {Pacucci} F.,  {Ferrara} A.,  {Agarwal} B.,  {Ricarte} A.,
  {Zackrisson} E.,   {Cappelluti} N.,  2017, \mn@doi [\apj]
  {10.3847/1538-4357/aa6330}, \href
  {https://ui.adsabs.harvard.edu/abs/2017ApJ...838..117N} {838, 117}

\bibitem[\protect\citeauthoryear{{Onoue} et~al.,}{{Onoue}
  et~al.}{2022}]{2022arXiv220907325O}
{Onoue} M.,  et~al., 2022, arXiv e-prints, \href
  {https://ui.adsabs.harvard.edu/abs/2022arXiv220907325O} {p. arXiv:2209.07325}

\bibitem[\protect\citeauthoryear{{Pacucci}, {Ferrara}, {Grazian}, {Fiore},
  {Giallongo}  \& {Puccetti}}{{Pacucci}
  et~al.}{2016}]{2016MNRAS_directcollapse}
{Pacucci} F.,  {Ferrara} A.,  {Grazian} A.,  {Fiore} F.,  {Giallongo} E.,
  {Puccetti} S.,  2016, \mn@doi [\mnras] {10.1093/mnras/stw725}, \href
  {https://ui.adsabs.harvard.edu/abs/2016MNRAS.459.1432P} {459, 1432}

\bibitem[\protect\citeauthoryear{{Peng}, {Ho}, {Impey}  \& {Rix}}{{Peng}
  et~al.}{2002}]{Galfit1}
{Peng} C.~Y.,  {Ho} L.~C.,  {Impey} C.~D.,   {Rix} H.-W.,  2002, \mn@doi [\aj]
  {10.1086/340952}, \href
  {https://ui.adsabs.harvard.edu/abs/2002AJ....124..266P} {124, 266}

\bibitem[\protect\citeauthoryear{{Peng}, {Ho}, {Impey}  \& {Rix}}{{Peng}
  et~al.}{2010}]{Galfit2}
{Peng} C.~Y.,  {Ho} L.~C.,  {Impey} C.~D.,   {Rix} H.-W.,  2010, \mn@doi [\aj]
  {10.1088/0004-6256/139/6/2097}, \href
  {https://ui.adsabs.harvard.edu/abs/2010AJ....139.2097P} {139, 2097}

\bibitem[\protect\citeauthoryear{{Perrin}, {Soummer}, {Elliott}, {Lallo}  \&
  {Sivaramakrishnan}}{{Perrin} et~al.}{2012}]{WebbPSF}
{Perrin} M.~D.,  {Soummer} R.,  {Elliott} E.~M.,  {Lallo} M.~D.,
  {Sivaramakrishnan} A.,  2012, in {Clampin} M.~C.,  {Fazio} G.~G.,  {MacEwen}
  H.~A.,   {Oschmann} Jacobus~M. J.,  eds,  Society of Photo-Optical
  Instrumentation Engineers (SPIE) Conference Series Vol. 8442, Space
  Telescopes and Instrumentation 2012: Optical, Infrared, and Millimeter Wave.
  p. 84423D, \mn@doi{10.1117/12.925230}

\bibitem[\protect\citeauthoryear{Pezzulli, Valiante  \& Schneider}{Pezzulli
  et~al.}{2016}]{stellarbh_mnras}
Pezzulli E.,  Valiante R.,   Schneider R.,  2016, \mn@doi [\mnras]
  {10.1093/mnras/stw505}, 458, 3047

\bibitem[\protect\citeauthoryear{{Pontoppidan} et~al.,}{{Pontoppidan}
  et~al.}{2022}]{Pontoppidan2022}
{Pontoppidan} K.,  et~al., 2022, arXiv e-prints, \href
  {https://ui.adsabs.harvard.edu/abs/2022arXiv220713067P} {p. arXiv:2207.13067}

\bibitem[\protect\citeauthoryear{{Prevot, M. L., et al.}}{{Prevot, M. L., et
  al.}}{1984}]{SMC}
{Prevot, M. L., et al.} 1984, \aap, \href
  {https://ui.adsabs.harvard.edu/abs/1984A&A...132..389P} {132, 389}

\bibitem[\protect\citeauthoryear{{Rieke} et~al.,}{{Rieke}
  et~al.}{2022}]{NirCam_paper}
{Rieke} M.~J.,  et~al., 2022, \mn@doi [arXiv e-prints]
  {10.48550/arXiv.2212.12069}, \href
  {https://ui.adsabs.harvard.edu/abs/2022arXiv221212069R} {p. arXiv:2212.12069}

\bibitem[\protect\citeauthoryear{{Rodighiero}, {Bisigello}, {Iani}, {Marasco},
  {Grazian}, {Sinigaglia}, {Cassata}  \& {Gruppioni}}{{Rodighiero}
  et~al.}{2023}]{2023MNRAS.518L..19R}
{Rodighiero} G.,  {Bisigello} L.,  {Iani} E.,  {Marasco} A.,  {Grazian} A.,
  {Sinigaglia} F.,  {Cassata} P.,   {Gruppioni} C.,  2023, \mn@doi [\mnras]
  {10.1093/mnrasl/slac115}, \href
  {https://ui.adsabs.harvard.edu/abs/2023MNRAS.518L..19R} {518, L19}

\bibitem[\protect\citeauthoryear{{Salpeter}}{{Salpeter}}{1955}]{Salpeter-IMF}
{Salpeter} E.~E.,  1955, \mn@doi [\apj] {10.1086/145971}, \href
  {https://ui.adsabs.harvard.edu/abs/1955ApJ...121..161S} {121, 161}

\bibitem[\protect\citeauthoryear{{Sarmento}, {Scannapieco}  \&
  {Pan}}{{Sarmento} et~al.}{2017}]{Sarmento2017}
{Sarmento} R.,  {Scannapieco} E.,   {Pan} L.,  2017, \mn@doi [\apj]
  {10.3847/1538-4357/834/1/23}, \href
  {https://ui.adsabs.harvard.edu/abs/2017ApJ...834...23S} {834, 23}

\bibitem[\protect\citeauthoryear{{Stalevski}, {Ricci}, {Ueda}, {Lira}, {Fritz}
  \& {Baes}}{{Stalevski} et~al.}{2016}]{Skirtor_mods}
{Stalevski} M.,  {Ricci} C.,  {Ueda} Y.,  {Lira} P.,  {Fritz} J.,   {Baes} M.,
  2016, \mn@doi [\mnras] {10.1093/mnras/stw444}, \href
  {https://ui.adsabs.harvard.edu/abs/2016MNRAS.458.2288S} {458, 2288}

\bibitem[\protect\citeauthoryear{{Thomas}, {Groves}, {Sutherland}, {Dopita},
  {Kewley}  \& {Jin}}{{Thomas} et~al.}{2016}]{Thomas2016}
{Thomas} A.~D.,  {Groves} B.~A.,  {Sutherland} R.~S.,  {Dopita} M.~A.,
  {Kewley} L.~J.,   {Jin} C.,  2016, \mn@doi [\apj]
  {10.3847/1538-4357/833/2/266}, \href
  {https://ui.adsabs.harvard.edu/abs/2016ApJ...833..266T} {833, 266}

\bibitem[\protect\citeauthoryear{{Trenti}, {Stiavelli}  \& {Shull}}{{Trenti}
  et~al.}{2009}]{late_pop_3}
{Trenti} M.,  {Stiavelli} M.,   {Shull} J.~M.,  2009, \mn@doi [\apj]
  {10.1088/0004-637X/700/2/1672}, \href
  {https://ui.adsabs.harvard.edu/abs/2009ApJ...700.1672T} {700, 1672}

\bibitem[\protect\citeauthoryear{{Treu} et~al.,}{{Treu}
  et~al.}{2022a}]{Treu2022}
{Treu} T.,  et~al., 2022a, arXiv e-prints, \href
  {https://ui.adsabs.harvard.edu/abs/2022arXiv220607978T} {p. arXiv:2206.07978}

\bibitem[\protect\citeauthoryear{{Treu} et~al.,}{{Treu}
  et~al.}{2022b}]{2022arXiv220713527T}
{Treu} T.,  et~al., 2022b, arXiv e-prints, \href
  {https://ui.adsabs.harvard.edu/abs/2022arXiv220713527T} {p. arXiv:2207.13527}

\bibitem[\protect\citeauthoryear{Trinca, Schneider, Maiolino, Valiante,
  Graziani  \& Volonteri}{Trinca et~al.}{2022}]{JWST_bhsearch}
Trinca A.,  Schneider R.,  Maiolino R.,  Valiante R.,  Graziani L.,   Volonteri
  M.,  2022, Seeking the growth of the first black hole seeds with JWST,
  \mn@doi{10.48550/ARXIV.2211.01389}, \url {https://arxiv.org/abs/2211.01389}

\bibitem[\protect\citeauthoryear{{Volonteri}, {Haardt}  \& {Madau}}{{Volonteri}
  et~al.}{2003}]{ApJ_smbh}
{Volonteri} M.,  {Haardt} F.,   {Madau} P.,  2003, \mn@doi [\apj]
  {10.1086/344675}, \href
  {https://ui.adsabs.harvard.edu/abs/2003ApJ...582..559V} {582, 559}

\bibitem[\protect\citeauthoryear{{Volonteri}, {Habouzit}  \&
  {Colpi}}{{Volonteri} et~al.}{2021}]{Volonteri2021}
{Volonteri} M.,  {Habouzit} M.,   {Colpi} M.,  2021, \mn@doi [Nature Reviews
  Physics] {10.1038/s42254-021-00364-9}, \href
  {https://ui.adsabs.harvard.edu/abs/2021NatRP...3..732V} {3, 732}

\bibitem[\protect\citeauthoryear{{Volonteri et al.}}{{Volonteri et
  al.}}{2023}]{AGN_SED_param}
{Volonteri et al.} 2023, \mn@doi [\mnras] {10.1093/mnras/stad499}, \href
  {https://ui.adsabs.harvard.edu/abs/2023MNRAS.521..241V} {521, 241}

\bibitem[\protect\citeauthoryear{{Windhorst} et~al.,}{{Windhorst}
  et~al.}{2023}]{Windhorst2023}
{Windhorst} R.~A.,  et~al., 2023, \mn@doi [\aj] {10.3847/1538-3881/aca163},
  \href {https://ui.adsabs.harvard.edu/abs/2023AJ....165...13W} {165, 13}

\bibitem[\protect\citeauthoryear{{Yang} et~al.,}{{Yang} et~al.}{2020a}]{Cigale}
{Yang} G.,  et~al., 2020a, \mn@doi [\mnras] {10.1093/mnras/stz3001}, \href
  {https://ui.adsabs.harvard.edu/abs/2020MNRAS.491..740Y} {491, 740}

\bibitem[\protect\citeauthoryear{{Yang} et~al.,}{{Yang}
  et~al.}{2020b}]{z75_blackhole}
{Yang} J.,  et~al., 2020b, \mn@doi [\apjl] {10.3847/2041-8213/ab9c26}, \href
  {https://ui.adsabs.harvard.edu/abs/2020ApJ...897L..14Y} {897, L14}

\bibitem[\protect\citeauthoryear{Yue, Ferrara, Salvaterra, Xu  \& Chen}{Yue
  et~al.}{2014}]{mnras_collapse_era}
Yue B.,  Ferrara A.,  Salvaterra R.,  Xu Y.,   Chen X.,  2014, \mn@doi [\mnras]
  {10.1093/mnras/stu351}, 440, 1263

\bibitem[\protect\citeauthoryear{{Zavala} et~al.,}{{Zavala}
  et~al.}{2023}]{Zavala2023}
{Zavala} J.~A.,  et~al., 2023, \mn@doi [\apjl] {10.3847/2041-8213/acacfe},
  \href {https://ui.adsabs.harvard.edu/abs/2023ApJ...943L...9Z} {943, L9}

\makeatother
\end{thebibliography}



\bsp	
\label{lastpage}
\end{document}